\documentclass[twocolumn,times]{aastex631}
\usepackage{tikz}
\usepackage{amsmath}
\usepackage{xcolor}
\usepackage{subfigure}
\usetikzlibrary{positioning}

\begin{document}

\title{Extracting the Epoch of Reionization Signal with 3D U-Net Neural Networks Using Data-driven Systematic Effect Model}

\author[0000-0001-5469-5408]{Li-Yang Gao}
\affiliation{Liaoning Key Laboratory of Cosmology and Astrophysics, College of Sciences, Northeastern University, Shenyang 110819, China}%
\affiliation{Kapteyn Astronomical Institute, University of Groningen, P.O. Box 800, 9700 AV Groningen, The Netherlands}%

\author[0000-0003-1840-0312]{L\'eon V. E. Koopmans}
\affiliation{Kapteyn Astronomical Institute, University of Groningen, P.O. Box 800, 9700 AV Groningen, The Netherlands}%

\author[0000-0003-3802-4289]{Florent G. Mertens}
\affiliation{Kapteyn Astronomical Institute, University of Groningen, P.O. Box 800, 9700 AV Groningen, The Netherlands}
\affiliation{LERMA, Observatoire de Paris, Universit\'e PSL, CNRS, Sorbonne Universit\'e, F-75014 Paris, France}

\author[0000-0001-9919-4121]{Satyapan Munshi}
\affiliation{Kapteyn Astronomical Institute, University of Groningen, P.O. Box 800, 9700 AV Groningen, The Netherlands}%

\author[0000-0003-1962-2013]{Yichao Li}
\affiliation{Liaoning Key Laboratory of Cosmology and Astrophysics, College of Sciences, Northeastern University, Shenyang 110819, China}%

\author[0000-0001-7507-6948]{Stefanie A. Brackenhoff}
\affiliation{Kapteyn Astronomical Institute, University of Groningen, P.O. Box 800, 9700 AV Groningen, The Netherlands}%

\author[0000-0002-3351-5778]{Emilio Ceccotti}
\affiliation{Kapteyn Astronomical Institute, University of Groningen, P.O. Box 800, 9700 AV Groningen, The Netherlands}%
\affiliation{INAF -- Istituto di Radioastronomia, Via P. Gobetti 101, 40129 Bologna, Italy}%

\author[0000-0002-1292-5268]{J. Kariuki Chege}
\affiliation{Kapteyn Astronomical Institute, University of Groningen, P.O. Box 800, 9700 AV Groningen, The Netherlands}%

\author[0000-0003-3401-4884]{Anshuman Acharya}
\affiliation{Max-Planck-Institut f\"{u}r Astrophysik, Garching 85748,
Germany}%

\author[0000-0001-9816-5070]{Raghunath Ghara}
\affiliation{Department of Physical Sciences, Indian Institute of Science Education and Research Kolkata, Mohanpur, WB 741246, India
}%

\author[0000-0002-2560-536X]{Sambit K. Giri}
\affiliation{Van Swinderen Institute for Particle Physics and Gravity, University of Groningen, Nijenborgh 4, 9747 AG Groningen, The Netherlands
}%

\author[0000-0002-5174-1365]{Ilian T. Iliev}
\affiliation{Department of Physics \& Astronomy, University of Sussex,  Brighton, BN1 9QH, UK
}%

\author[0000-0002-2512-6748]{Garrelt Mellema}
\affiliation{Dept. of Astronomy \& Oskar Klein Centre, Stockholm University, AlbaNova, Stockholm University, SE-106 91 Stockholm, Sweden
}%

\author[0000-0002-6029-1933]{Xin Zhang}
\correspondingauthor{Xin Zhang}
\email{zhangxin@mail.neu.edu.cn}
\affiliation{Liaoning Key Laboratory of Cosmology and Astrophysics, College of Sciences, Northeastern University, Shenyang 110819, China}%
\affiliation{MOE Key Laboratory of Data Analytics and Optimization for Smart Industry,
Northeastern University, Shenyang 110819, China}%
\affiliation{National Frontiers Science Center for Industrial Intelligence and Systems Optimization, Northeastern University, Shenyang 110819, China}%

\begin{abstract}

Neutral hydrogen (HI) serves as a crucial probe for the Cosmic Dawn and the Epoch of Reionization (EoR). Actual observations of the 21-cm signal often encounter challenges such as thermal noise and various systematic effects. 
To overcome these challenges, we simulate SKA-Low-depth images in South Celestial Pole (SCP) field and process them with a deep learning method. 
We utilized foreground residuals acquired by LOFAR during actual North Celestial Pole (NCP) field observations, thermal and excess variances calculated via Gaussian process regression (GPR), and 21-cm signals generated with \textbf{\texttt{21cmFAST}} for signal extraction tests. 
Our approach to overcome these foreground, thermal noise, and excess variance components employs a 3D U-Net neural network architecture for image analysis. 
When considering thermal noise corresponding to 1752 hours of integration time, U-Net provides reliable 2D power spectrum predictions, and robustness tests ensure that we get realistic EoR signals. 
Adding foreground residuals, however, causes inconsistencies below the horizon delay-line. 
Lastly, evaluating both thermal noise and excess variances with observations up to 4380 hours and 13140 hours ensures reliable power spectrum estimations within the EoR window and across nearly all scales, respectively. The incoherence of excess variances in the frequency direction can greatly affect deep learning to extract 21-cm signals.
\end{abstract}

\keywords{HI line emission(690) --- Reionization(1383) --- Gaussian Processes regression(1930) --- Neural networks(1933)}

\section{Introduction}
\label{Introduction}
To fully explore the Cosmic Dawn \citep[CD,][]{Pritchard:2006sq} ($12 < z < 30$) and the Epoch
of Reionization \citep[EoR,][]{Madau_1997} ($6 < z < 12$), there is a need for a new probe of the infant Universe, beyond infrared observations with the James Webb Space Telescope (JWST), Hubble Space Telescope (HST), and mm/sub-mm observations with the Atacama Large Millimeter/submillimeter Array (ALMA), which only probe the brightest galaxies at these early epochs.
The 21-cm signal is regarded as the most promising probe for detecting the distribution of neutral hydrogen (HI) in the inter-galactic medium (IGM) during the CD and the EoR \citep{Madau_1997, Shaver:1999gb, Furlanetto:2006jb, Pritchard_2012, Zaroubi2013}, which can help us understand the formation of the first generation of stars as well as galaxies and the evolution of the infant Universe.

Currently, there are several 21-cm experiments targeting at mapping the 21-cm power spectrum 
of EoR/CD, e.g., 
The 21 CentiMeter Array \footnote{\url{https://english.nao.cas.cn/}}
\citep[21CMA,][]{2009AAS...21322605W},
The Giant Metrewave Radio Telescope\footnote{\url{https://www.gmrt.org/}}
\citep[GMRT,][]{ananthakrishnan1995giant} EoR experiment, 
the Murchison Widefield Array\footnote{\url{https://www.mwatelescope.org/}} 
\citep[MWA,][]{Barry:2019qxp, Li:2019kqp, Trott:2020szf},
the Owens Valley Radio Observatory - Long Wavelength Array
\footnote{\url{https://www.ovro.caltech.edu/}} \citep[OVRO-LWA,][]{Eastwood:2019rwh},
the Low-Frequency Array\footnote{\url{http://www.lofar.org/}},
\citep[LOFAR,][]{van2013lofar,10.1093/mnras/stz1937,10.1093/mnras/staa3093}, 
the New Extension in Nan\c{c}ay Upgrading LOFAR\footnote{\url{https://nenufar.obs-nancay.fr/en/homepage-en/}} (NenuFAR),
the Hydrogen Epoch of Reionization Array\footnote{\url{https://reionization.org/}} 
\citep[HERA,][]{DeBoer:2016tnn}, etc.

The EoR/CD experiments of the current generation are mostly sensitivity-limited. 
A couple of the 21-cm power spectrum upper-limit constraints are reported.
The GMRT EoR experiment reported the 21-cm power spectrum upper limit of $\Delta^2_{21} < (248~\rm{mK})^2$ at 
$k = 0.50~h\,\rm{cMpc}^{-1}$ and $z \approx 8.6$ \citep{Paciga:2013fj}.
\cite{Yoshiura:2021yfx} reported $\Delta^2_{21} < 6.3 \times 10^6\,\rm{mK}^2$ 
at $k = 0.14~h\,\rm{cMpc}^{-1}$, $z \approx 15.2$ using 
$5.5$ hours observation data of the MWA.
\cite{Mertens:2020llj} achieved a 2$\sigma$ upper limit of $\Delta^2_{21} < (73~\rm{mK})^2$ at $k = 0.075~h\,\rm{cMpc}^{-1}$, $z \approx 9.1$ 
based on $141$ hours of LOFAR observations of the North Celestial Pole (NCP) field
\citep{Yatawatta:2013im, Patil:2017zqk}.
\cite{Garsden:2021kdo} reported an upper limit of $\Delta^2_{21} < 2 \times 10^{12} \rm{mK}^2$ at $k = 0.3~h\,\rm{cMpc}^{-1}$
with the median redshift of $z = 28$ 
using a $4$-hour observation from the OVRO-LWA.
\cite{Munshi:2023buw} reports a 2$\sigma$ upper limit of $\Delta^2_{21} < 2.4 \times 10^7~\rm{mK}^2$ at $k = 0.041~h\,\rm{cMpc}^{-1}$ and $z = 20.3$ 
using one-night observation of NenuFAR. 
The next generation EoR/CD experiments began to collect observation data and 
publish their early scientific results.
HERA Phase I gave their early constraints on the 21-cm power spectrum
2 $\sigma$ upper limits with $36$-hour observation 
-- $\Delta^2_{21} < (30.76)^2~\rm{mK}^2$ at $k = 0.192~h\,\rm{cMpc}^{-1}$, $z = 7.9$
and $\Delta^2_{21} < (95.74)^2~\rm{mK}^2$ at $k = 0.256~h\,\rm{cMpc}^{-1}$, $z = 10.4$, respectively. 
For the next generation EoR/CD experiments, e.g. the
HERA Phase II and Square Kilometer Array Low-frequency array
\footnote{\url{https://www.skao.int/}} \citep[SKA-Low,][]{Koopmans:2015sua}, 
are expected to have higher sensitivity and thus significantly improve the
constraints on the 21-cm power spectrum. 

However, there are still many difficulties in extracting the 21-cm brightness 
fluctuations of the EoR/CD.
The brightness temperature of the foreground contamination components 
are orders of magnitude higher than the  21-cm brightness fluctuation, which 
is known as the major challenge not only for EoR/CD studies but also for 
the HI intensity mapping analysis in the post-EoR Universe
\citep{2021MNRAS.504..208C, Spinelli:2021emp, Wolz:2015sqa}. 
Due to the smooth frequency dependence of the foregrounds, there are ways to separate them from the faint 21-cm signal \citep{Jelic:2008jg, 2009ApJ...695..183B, Ansari:2011bv, 2012MNRAS.423.2518C}.
However, the foreground is affected by chromatic instrumental effects, such as beam effects \citep{Asad:2015sda} and polarization leakage \citep{Asad:2015sda, Nunhokee:2017aaj, Bhatnagar:2001mh}.
Thus, the model-independent foreground subtraction methods, such as
principal component analysis \citep[PCA,][]{Masui:2012zc}, fast independent component analysis \citep[\textsc{FastICA},][]{761722,Chapman:2012yj,Wolz:2013wna,Cunnington:2019lvb}, and generalized morphological component analysis \citep[GMCA,][]{Patil:2014dpa, 4337755}, etc,
are mostly used in real data analysis.
 
The model-independent foreground subtraction methods may lead to either foreground residual or 
significant signal loss.
To overcome the challenges, we propose to eliminate the systematic effects using a deep learning approach.
Inspired by \cite{Makinen:2020gvh}, we used the 3D U-Net architecture to deal with the polarization leakage \citep{Gao:2022xdb} and the beam effect \citep{Ni:2022kxn} 
for the post-EoR HI intensity mapping survey.

However, such supervised deep-learning methods are penitentially model-dependent and 
the prior knowledge of the systematic model impacts the results \citep{2024MNRAS.532.2615C}.
In this work, we improve the deep-learning-based systematic effect elimination algorithm,
using the training set generated with Gaussian Process Regression (GPR) in the public code \textbf{\texttt{ps\_eor}}\footnote{\url{https://gitlab.com/flomertens/ps_eor/}},
which injects the systematic effects according to real observation data.

This paper is organized as follows.
We simulated the EoR signals and the thermal noises in Section~\ref{Simulation} and modeled the systematic effects in Section~\ref{systematic_effects}, the deep learning methodology in Section~\ref{Method}, the results along with the discussions in Section~\ref{Results_and_discussion}, and the conclusion in Section~\ref{Conclusion}.

\section{SKA-Low EoR simulation}
\label{Simulation}

\begin{figure*}[htp]
    \centering
    \subfigure{\includegraphics[width=0.45\textwidth]{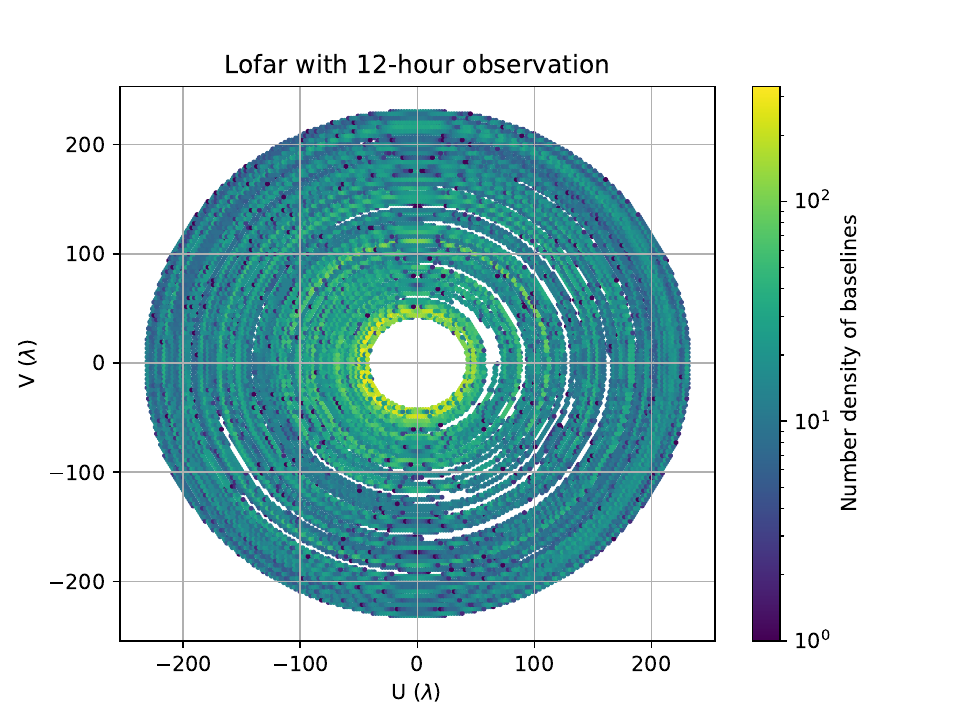}}
    \subfigure{\includegraphics[width=0.45\textwidth]{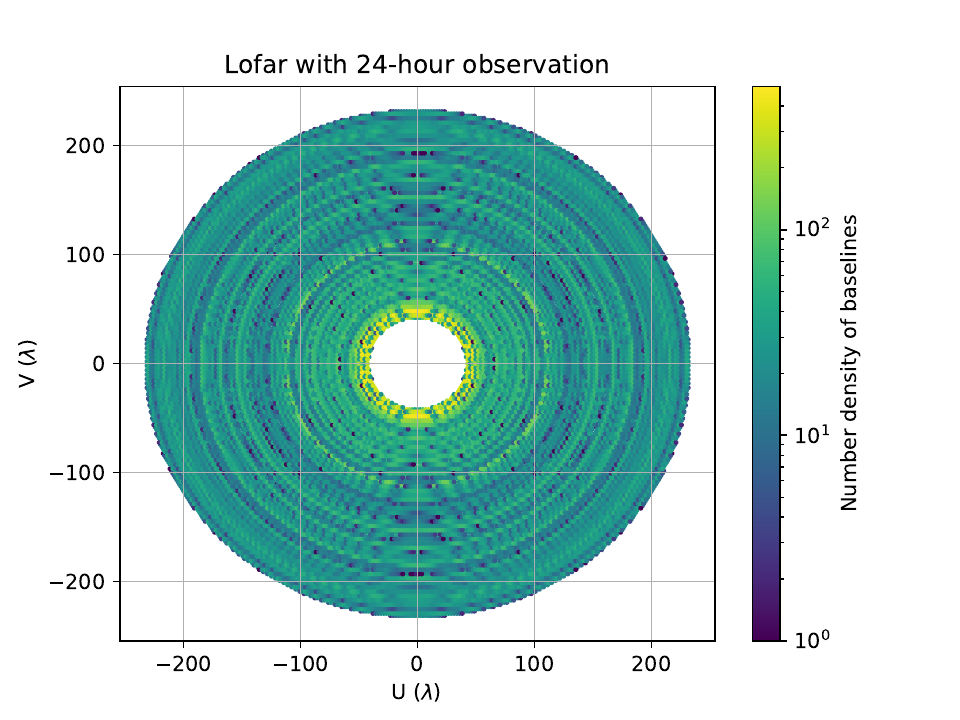}}\\
    \subfigure{\includegraphics[width=0.45\textwidth]{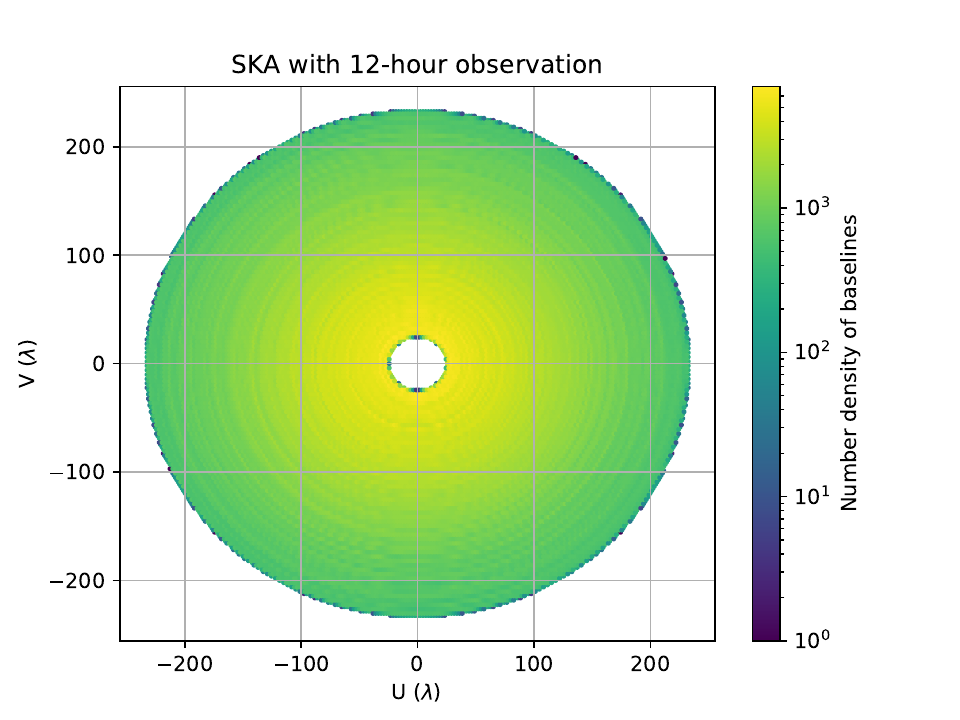}}
    \subfigure{\includegraphics[width=0.45\textwidth]{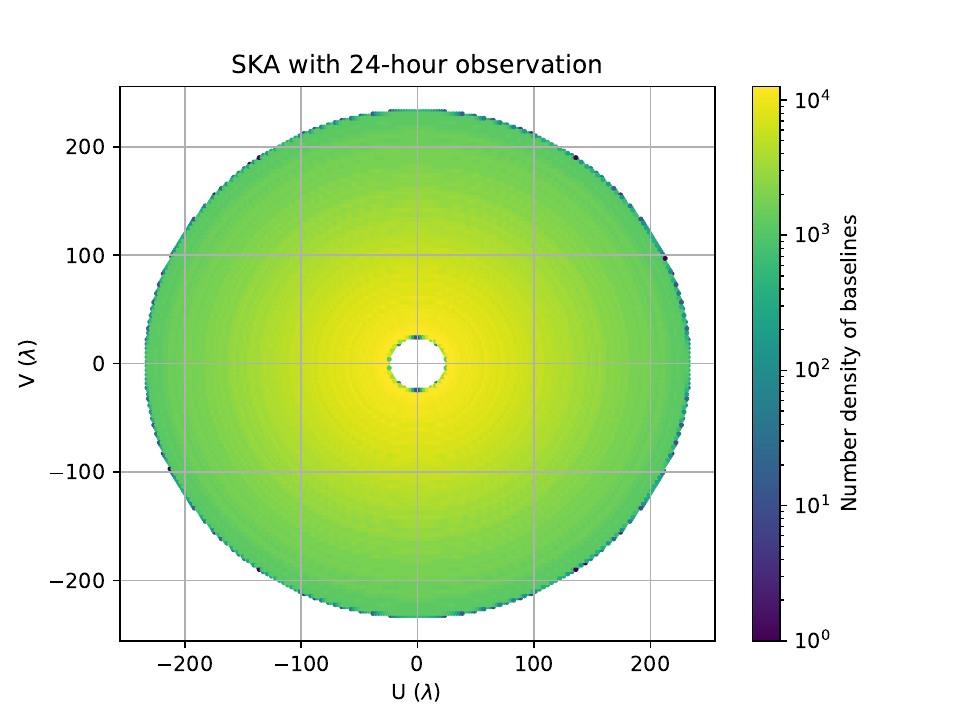}}
    \caption{\label{number_density}Number densities of LOFAR baselines (first line) and SKA baselines (second line) with 12-hour observation (first column) and 24-hour observation (second column) per day.}
\end{figure*}

The main focus of this work is on the impact of systematic effects on observations in the CP (Celestial Pole) field, i.e., the NCP field for LOFAR and the South Celestial Pole (SCP) field for SKA.
We simulated the baseline number densities for various observation times using the built-in baseline configurations of LOFAR and SKA-Low provided by \textbf{\texttt{ps\_eor}}.
The number densities of the baselines for 12 and 24 hours of continuous observation times in one day are shown in Fig.~\ref{number_density}.
The number densities of baselines and the $uv$-coverage of SKA are significantly better than those of LOFAR. 
SKA's 12-hour observations provide perfect coverage of the entire observation area, whereas LOFAR's 12-hour observations result in some gaps in $uv$-coverage.
Although a longer observation time leads to a greater baseline density, there are many special circumstances that prevent real observations from being made.
During the first three cycles of the LOFAR EoR Key Science Project, a total of 141 hours of observational data 
were collected over 10 days, averaging 14 hours per day. 
An optimized observation strategy should account for variable radio contamination sources, 
such as solar radiation, ionospheric conditions, and human activity.
For these reasons, we use 12 hours of observations per day for the SKA simulations.
Since too many systematic effects may require a significantly larger neural network architecture, we ignore the effect of gains from different directions on the simulation. 
Considering that this study evaluates multiple components identified by distinct subscripts, Table~\ref{variables} within Appendix~\ref{appendixA} catalogs these subscripts to enhance clarity.

\subsection{EoR 21-cm signal}
\label{EoR_signal}

\begin{figure}
\centering
\includegraphics[width=0.50\textwidth]{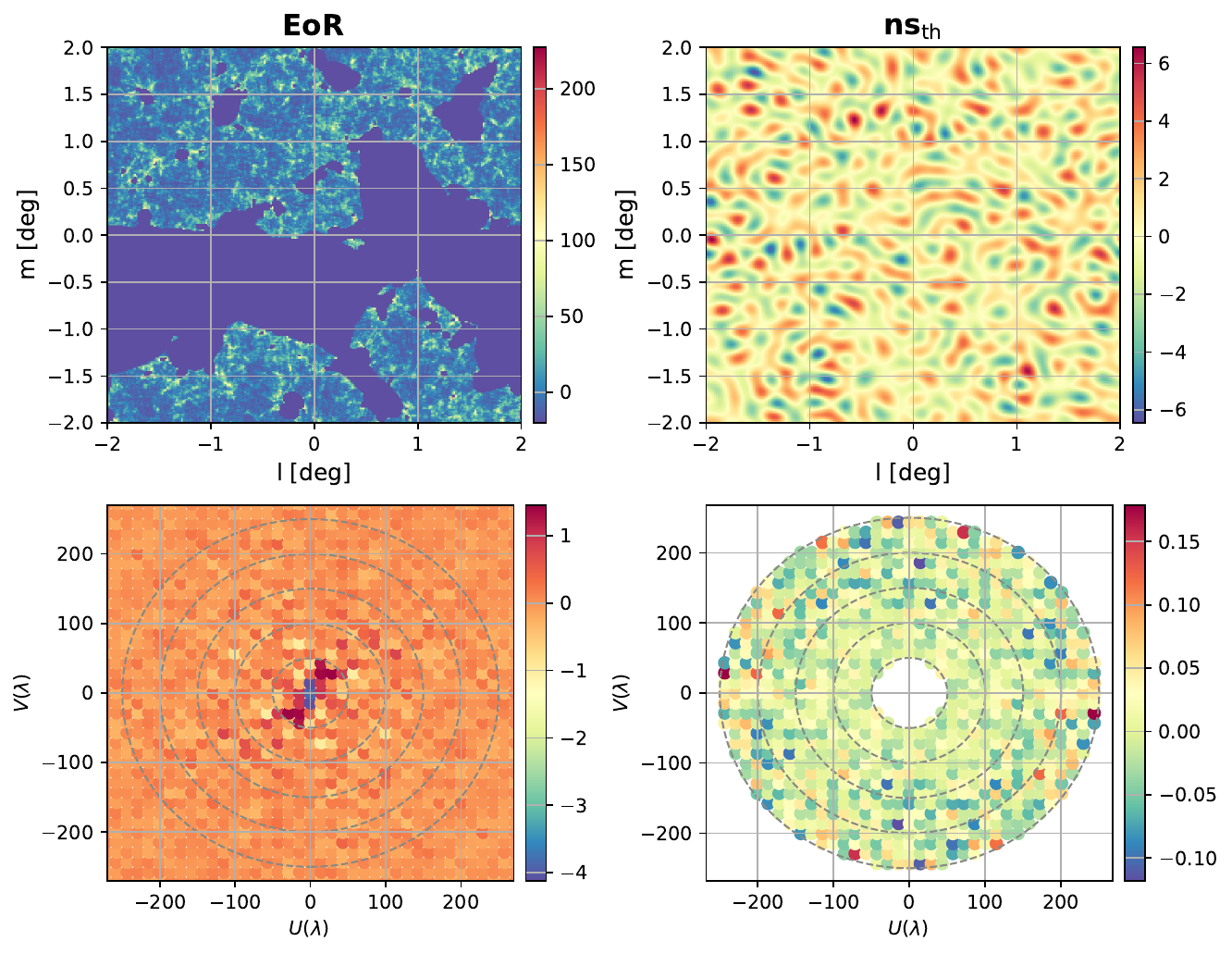}
\caption{\label{th_eor_component_img_visi}Simulated slices of images (first line) and gridded visibilities (second line) from SKA before applying LOFAR $uv$-coverage of $\mathbf{EoR}$ and $\mathbf{ns}_\text{th}$, where $\mathbf{ns}_\text{th}$ is obtained based on LOFAR's imaging capabilities and sensitivity, and $\mathbf{EoR}$ is obtained from full $uv$-coverage using \textbf{\texttt{21cmFAST}} code. The units of images and gridded visibilities are both millikelvin (mK).}
\end{figure}


We employ the public code \textbf{\texttt{21cmFAST}}
\citep{10.1111/j.1365-2966.2010.17731.x, Murray:2020trn} to simulate the EoR 21-cm brightness 
temperature cubes for the SKA-Low observations.
We set the cosmological parameters $\{h, \Omega_{\rm m}, \Omega_{\rm b},T_{\rm cmb}, n_s, \sigma_8\}$ to $\{0.676, 0.31, 0.049,2.725~{\rm K}, 0.9665, 0.8\}$. The ionizing efficiency of high-$z$ galaxies $\zeta$ is set to 30,
and other relevant astrophysical parameters are set to default values.
The cube corresponds to our redshift of interest ($z \approx 9.1$) and contains a 
$4^\circ \times 4^\circ$ sky field with $256$ pixels in each direction. 
In order to be consistent with the number of frequency channels of the real LOFAR observation, we averaged every quartet of pixels along the line-of-sight axis. 
This process generated a voxel cube of dimensions $64\times256\times256$. 
We denote this component as $\mathbf{EoR}$.
Fig.~\ref{th_eor_component_img_visi} presents a frequency slice of the $\mathbf{EoR}$ cube 
(top panel) and the corresponding $uv$-coverage (bottom panel) in the left column.
The $uv$-coverage slice of EoR provided by \textbf{\texttt{21cmFAST}} is full coverage, 
resulting in a more pronounced structure compared to the other components.
It is clear that the HI distribution during the EoR is primarily shaped by the sizes and 
spatial arrangement of the ionized regions. 
When the comoving sizes of these ionized regions are on the scale of a few Mpc, 
the large-scale statistical properties of the HI distribution are predominantly influenced by 
the Poisson noise associated with the discrete ionized regions, 
resulting in a highly non-Gaussian distribution.

Considering the processing speed of GPR and the anticipated volume of datasets required 
for subsequent deep learning applications, we generated a total of $536$ datasets. 
Due to time constraints, we used the \textbf{\texttt{21cmFAST}} code to produce 
only 134 brightness temperature cubes for the $\mathbf{EoR}$. To augment the data, 
we rotated the cubes by $90^\circ$, $180^\circ$, and $270^\circ$ around the line of sight relative to the field center.
This augmentation process resulted in a total of $536$ enhanced data cubes for the $\mathbf{EoR}$.

\subsection{Thermal noise}
\label{Thermal_noise}


Based on the antenna configurations of SKA-Low and LOFAR, the frequency range of the observations, the orientation of the target sky area, and the size of the field of view (FOV), we modeled the System Equivalent Flux Densities (SEFDs) and capabilities of these two observations. 
In this work, we consider a $4\times 4\,{\deg}^2$ field with frequency range of $134 - 146$ MHz, 
which are consistent with the real LOFAR observations of the NCP field.
We assume the frequency channel width of 0.195 MHz and visibility data are collected every $1$ s.
The thermal noise simulations are also carried out with the \textbf{\texttt{ps\_eor}}.
Note that the $uv$-coverage is restricted to baselines spanning $50-250 \lambda$  to be consistent with the real LOFAR observations.
The thermal noise is denoted as $\mathbf{ns}_\text{th}$ and the corresponding image and $uv$-coverage assumed for SKA are shown in Fig.~\ref{th_eor_component_img_visi} in the right column along with the $\mathbf{EoR}$ component.

We evaluated the amplitude variance between two simulated thermal noise images, as shown in Fig.~\ref{noise_image}.
It is observed that the thermal noise after 1752 hours of SKA-Low observations is approximately $5\%$ of the thermal noise present in current LOFAR observations with the same integration time.
Meanwhile, in order to compare the effect of different integration times on the extraction of the 21-cm signals, we similarly simulated the noises with integration times of 4380 hours and 13140 hours.
The durations of 1752, 4380, and 13140 hours represent 12 hours of daily observations conducted over periods of 0.4, 1, and 3 years, respectively.
Corresponding results are shown in Section \ref{differtime}.

\section{Data-driven systematic effects modeling}
\label{systematic_effects}

Extracting the 21-cm signal remains highly challenging due to the presence of numerous systematic 
effects that cannot be fully subtracted from current interferometric array observations. 
These include calibration errors, primary beam imperfections, 
$uv$-coverage limitations with bright source masking, 
radio frequency interference (RFI) masking, and other unidentified effects. 
Although systematic effects can be modeled, such approaches often yield incomplete 
or inaccurate representations. 
A systematic effects model based on real observations would potentially address these challenges.

We constructed a systematic effects model focused on a $4^\circ \times 4^\circ$ field around 
the CP field, leveraging observations from the 
LOFAR Epoch of Reionization (EoR) Key Science Project utilizing the High-Band Antenna (HBA) system.
The data comprise unsubtracted NCP observations collected over 141 hours during LOFAR 
Cycles 0, 1, 2, and 3, using all core stations in split mode (48 stations in total) 
alongside remote stations. 
The configuration of LOFAR stations defines the $uv$-coverage as a function of frequency, and our analysis focuses on the $134 - 146$ MHz frequency range (corresponding to a redshift interval of $z \approx 8.7 - 9.6$). The $uv$-coverage is restricted to baselines spanning $50-250 \lambda$. Notably, the flagging of residuals from CasA and CygA produced a cross-shaped pattern in the $uv$-coverage, as depicted in Fig.~\ref{fg_ex_component_img_visi}.

After processing through the LOFAR pipeline \citep{Mertens:2020llj}, we obtained foreground residuals comprising multiple components: residual smooth foregrounds, thermal noise, excess power from systematic effects, and the 21-cm signal. The resulting datacube from the observation contains $64\times480\times480$ voxels across 64 frequency channels. Due to the need to maintain shape consistency with the previous components and the computational demands of deep learning, including constraints on GPU memory and training times, we balanced network complexity and dataset size by down-sampling the datacube resolution to $64\times256\times256$. To further analyze the data, Gaussian process regression was employed to model the individual components.

\subsection{Gaussian process regression}
\label{Gaussian_Process_Regression}
Gaussian Process (GP) modeling is a non-parametric Bayesian method used to model functions
over a continuous input space \citep{10.7551/mitpress/3206.001.0001, Gelman2013BayesianDA}. 
A GP is defined as a collection of random variables or vectors, such that the joint distribution 
of any finite subset of these variables is a multivariate Gaussian distribution,
\begin{equation}
\mathbf{f}(\mathbf{x}) \sim \mathcal{GP}(m(\mathbf{x}), \mathbf{K}(\mathbf{x}, \mathbf{x}')),
\end{equation}
where $\mathbf{x}$ represents points in the input space,  $m(\mathbf{x})$  denotes the mean function, 
and $\mathbf{K}(\mathbf{x}, \mathbf{x}{\prime})$  is the covariance matrix, also referred to as the kernel. 
Using this formulation, the joint distribution of the random variables $\mathbf{f}(\mathbf{x})$  
can be obtained. 

After processing using the LOFAR-EoR pipeline \citep{Mertens:2020llj}, we obtain the gridded visibility data cube 
$\tilde{T}(u, v, \nu)$.
The data $\mathbf{d}$ can be viewed as the sum of multiple components, i.e., 
foreground $\mathbf{f}_{\text{fg}}$, excess variance $ \mathbf{f}_\text{ex}$, 
thermal noise $\mathbf{f}_\text{th}$, and 21-cm EoR signal $\mathbf{f}_\text{EoR}$,
\begin{equation}
\mathbf{d} = \mathbf{f}_{\text{fg}}(\nu) + \mathbf{f}_\text{ex}(\nu) + \mathbf{f}_\text{th}(\nu) + \mathbf{f}_\text{EoR}(\nu) .
\end{equation}
Each component is a function of frequency  $\nu$, and their distinct spectral behaviors 
allow them to be distinguished theoretically through the use of specific kernels in GPR. 
The covariance matrix for the data, $\mathbf{K}(\nu_p, \nu_q)$, is given by,
\begin{multline}
\mathbf{K}(\nu_p, \nu_q) = \mathbf{K}_{\text{fg}}(\nu_p, \nu_q) + \mathbf{K}_{\text{ex}}(\nu_p, \nu_q) \\
+ \mathbf{K}_{\text{th}}(\nu_p, \nu_q) + \mathbf{K}_{\text{EoR}}(\nu_p, \nu_q).
\label{k_pq}
\end{multline}

The foreground primarily arises from diffuse Galactic emissions and extragalactic point sources, 
while the noise originates from thermal emissions in antennas, receivers, and related instrumentation. 
Excess variance represents additional power with small coherence scales,
which is often associated with systematic effects and challenged to be distinguished from
the 21-cm signal.
The covariance matrix 
$\mathbf{K}_{\text{fg}}(\nu_p, \nu_q)$ and $\mathbf{K}_{\text{ex}}(\nu_p, \nu_q)$ 
are modeled using the Mat\'ern class kernels \citep{Stein}, 
\begin{equation}
\kappa_{\text{M}}(r) = \sigma^2 \frac{2^{1-\eta}}{\Gamma(\eta)} \left( \frac{\sqrt{2\eta} r}{l} \right)^\eta K_\eta \left( \frac{\sqrt{2\eta} r}{l} \right),
\label{martern}
\end{equation}
where $\sigma^2$ represents the variance, $\eta$ denotes the functional forms of 
Mat\'ern class kernels in special cases, $\Gamma$ is the Gamma function, 
$l$ represents the frequency coherence scale, 
$r = \lvert \nu_p - \nu_q\rvert$ indicates the frequency separation,
and $K_\eta$ refers to the modified Bessel function of the second kind.

Using Markov Chain Monte Carlo (MCMC) methods, we constrain the hyperparameters of these 
covariance matrices based on real LOFAR observations of the NCP field, 
enabling the isolation of cleaner components and reducing contamination.

\subsection{Foreground residual}
\label{Foreground_Residual}

\begin{figure}
\centering
\includegraphics[width=0.50\textwidth]{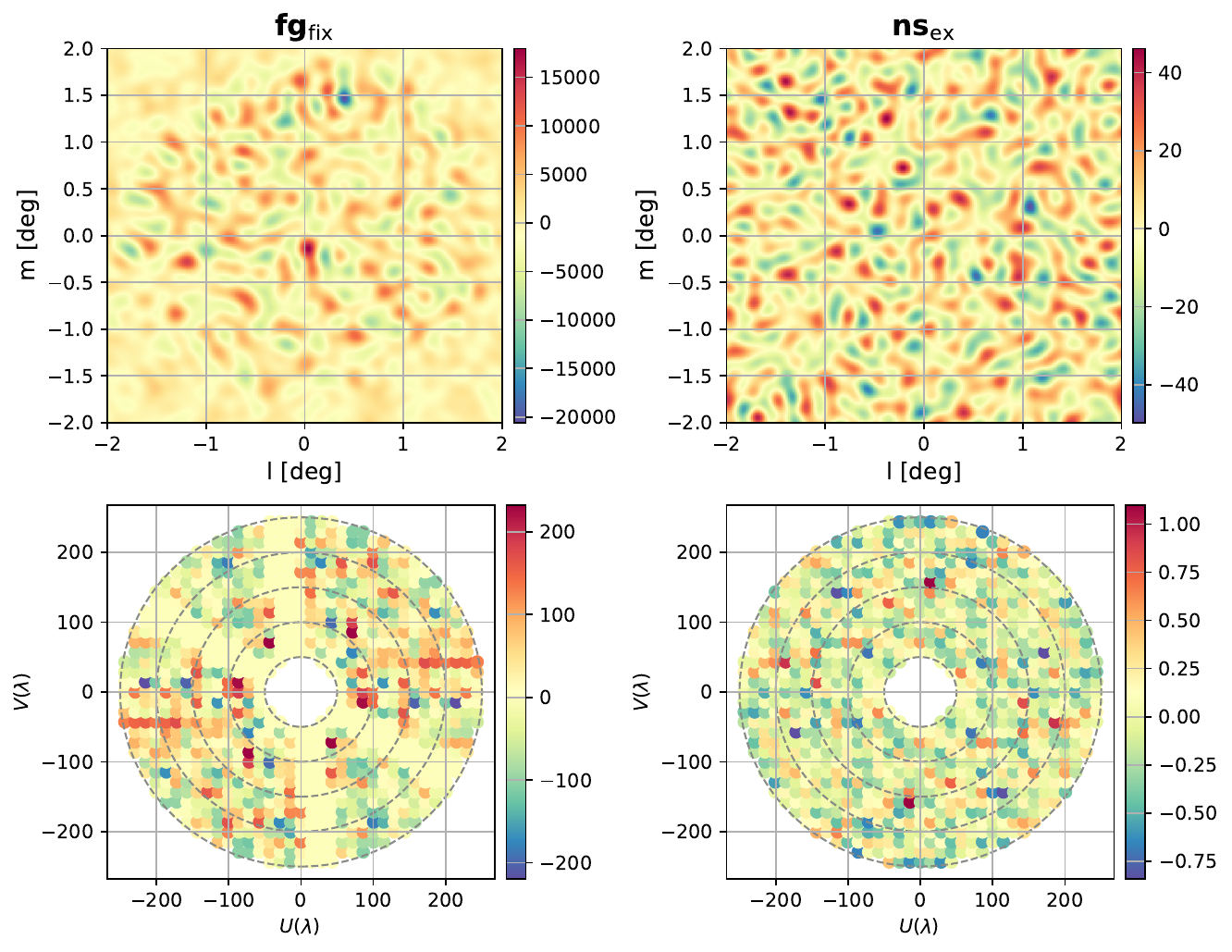}
\caption{\label{fg_ex_component_img_visi}Simulated slices of images (first line) and gridded visibilities (second line) from SKA before applying LOFAR $uv$-coverage of $\mathbf{fg}_\text{fix}$ and $\mathbf{ns}_\text{ex}$, where $\mathbf{fg}_\text{fix}$ is the smooth foreground residual obtained via GPR based on observations of the real NCP sky field,  $\mathbf{ns}_\text{ex}$ is obtained based on LOFAR's imaging capabilities, sensitivity, and $\mathbf{ns}_\text{th}$. The units of images and gridded visibilities are both mK.}
\end{figure}

We generate the foreground residual that we need based on $\mathbf{K}_{\text{fg}}(\nu_p, \nu_q)$. 
The foreground residual primarily consists of intrinsic sky emissions, 
including contributions from confusion-limited extragalactic sources and our Galaxy, characterized by 
$\mathbf{K}_\text{int}(l_\text{int}, \sigma^2_\text{int})$ \citep{Mertens:2017gxw}, 
as well as mode-mixing contaminants, represented by 
$\mathbf{K}_\text{mix}(l_\text{mix}, \sigma^2_\text{mix})$\citep{Morales:2012kf, Vedantham:2011mh}.

\begin{table}
\centering
\caption{The best-fit values from MCMC of Hyper-parameters corresponding to different kernels \citep{Mertens:2020llj}. The unit of $l$ is MHz and the unit of $\sigma^2$ is $\rm{mK}^2$.}
\begin{tabular}{llcc}
\hline
$\mathbf{K}_\eta$ & $\eta$& $l$&$\sigma^2$ \\
\hline
$\mathbf{K}_\text{int}$&$  $&$30$&$0$\\
$\mathbf{K}_\text{mix}$&$ 3/2$&$8.1$&$50.4 \sigma_n^2$ \\
$\mathbf{K}_\text{ex}$&$5/2$&$0.26$&$2.18 \sigma_n^2$ \\

\hline
\end{tabular}
\centering
\label{bestfit}
\end{table}

\begin{figure}
\centering
\includegraphics[width=0.50\textwidth]{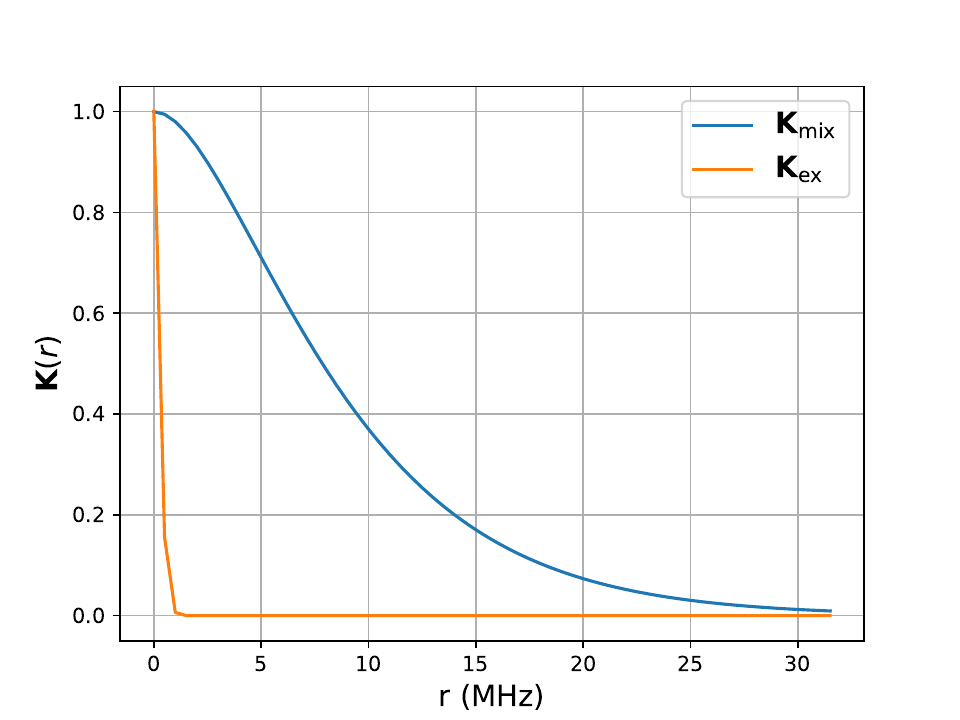}
\caption{\label{kr}Mat\'ern covariance functions for $\mathbf{K}_\text{mix}$ and $\mathbf{K}_\text{ex}$.}
\end{figure}

Based on the real observation data cube from LOFAR, we obtain the best-fit values of the hyperparameters in $\mathbf{K}_\text{int}(l_\text{int} , \sigma^2_\text{int})$ and $\mathbf{K}_\text{mix}(l_\text{mix}, \sigma^2_\text{mix})$ by MCMC and model the foreground residuals and the best-fit values of the hyperparameters are shown in Table~\ref{bestfit}. 
The variation of $\mathbf{K}_\text{mix}$ with scale is shown in Fig.~\ref{kr}.
We show this table only to compare the foreground and excess variance in terms of scale and variance.
The scale for $\mathbf{K}_\text{int}$ in LOFAR's real observations is about 30 MHz and for $\mathbf{K}_\text{mix}$ is about 8.1 MHz. 
Based on the constraint results of MCMC from \cite{Mertens:2020llj}, the variance of $\mathbf{K}_\text{mix}$ is set to $50.4 \sigma_n^2$.
Since the real foreground does not change, the smooth foreground residual in this part is kept fixed. 
Hence, random realizations of the residuals, during training of the network, arise only from the excess variance, thermal noise, and 21-cm signals, which will be elaborated on later. 
Since LOFAR and SKA have similar antenna placement strategies, we believe that they have statistically similar behavior in their observations. 
In this study, we specifically investigate the effects of $uv$-coverage and excess variance on the extraction of the 21-cm signal. 
Thus, the actual foreground residuals from LOFAR's NCP sky field are used as the mock foreground residuals for the SKA.
In Fig.~\ref{fg_ex_component_img_visi}, we show a slice of the $\mathbf{fg}_\text{fix}$ cube.

\subsection{Excess variance}
\label{Excess_noise}
\begin{figure}
\centering
\includegraphics[width=0.50\textwidth]{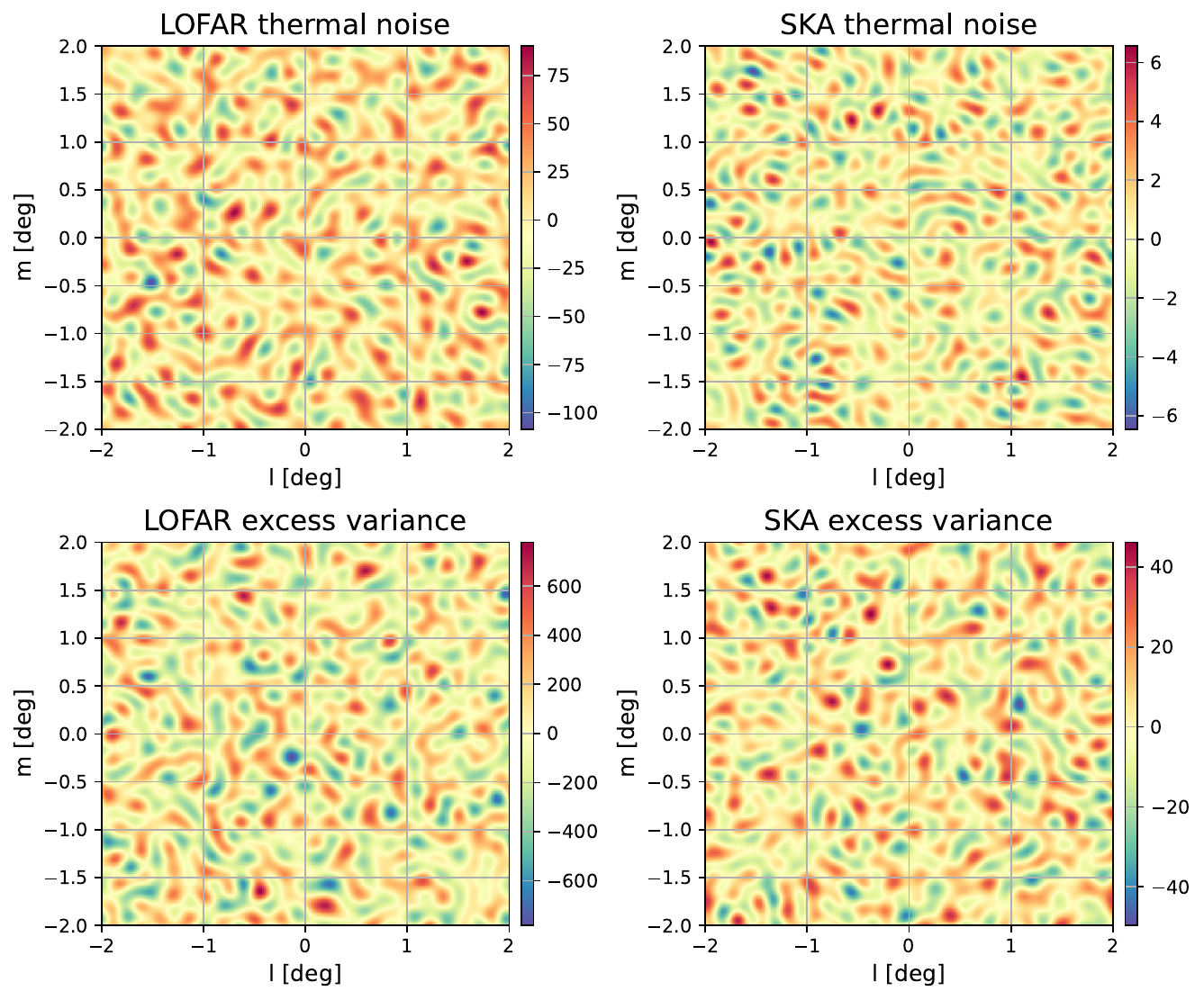}
\caption{\label{noise_image}
Thermal noise slices and excess variance slices for LOFAR and SKA, generated by GPR.
The unit of images is mK.}
\end{figure}

\cite{Mertens:2020llj} and \cite{Munshi:2023buw} demonstrate that the data contain extra power on a small coherence scale, with a strength between the foreground residual and the 21-cm signal. 
This additional power primarily stems from systematic errors, including instrumental effects, RFI, and suboptimal calibration.
Because of the typically small-scale nature of these excess components in the residuals, it is not easy to differentiate them from the 21-cm signal. 
Due to the complexity of modeling these components, we continue employing GPR for excess variance simulations to generate mocks.
An exponential covariance model with $\eta_\text{ex} = \frac{5}{2}$ is strongly favored by the observation data from LOFAR in Eq.~(\ref{martern}) and we finally get $\mathbf{K}_\text{ex}(l_\text{ex}, \sigma^2_\text{ex})$.
This component is denoted as $\mathbf{ns}_\text{ex}$, and the structure of the data cube for $\mathbf{ns}_\text{ex}$ is identical to that of $\mathbf{fg}_\text{fix}$.
The $141$-hour LOFAR observations of the NCP field \citep{Mertens:2020llj} reveal an excess variance with $l_\text{ex} = 0.26$ MHz. 
Furthermore, the excess variance $\sigma^2_\text{ex}$ is found to be $2.18$ times the thermal noise variance $\sigma^2_\text{ns}$ \citep{Mertens:2020llj} which can be seen in Table~\ref{bestfit}.
The variation of $\mathbf{K}_\text{ex}$ with scale is shown in Fig.~\ref{kr}.
Due to the very low correlation of the excess variance in frequency compared to the foreground, excess variance is very difficult to subtract.
Based on these assumptions, we construct excess variance cubes for LOFAR and SKA, using the thermal noise variances discussed in the previous subsection, as illustrated in Fig.~\ref{noise_image}, and assuming that the ratio between excess variance and thermal noise is invariant.
The latter is partly motivated by the fact that the excess variance appears to be largely incoherent between different observations \citep{Mertens:2020llj}.
To facilitate comparison, Fig.~\ref{fg_ex_component_img_visi} shows the excess variance image and $uv$-coverage of SKA alongside the other components.

\begin{figure*}
\centering
\includegraphics[width=0.95\textwidth]{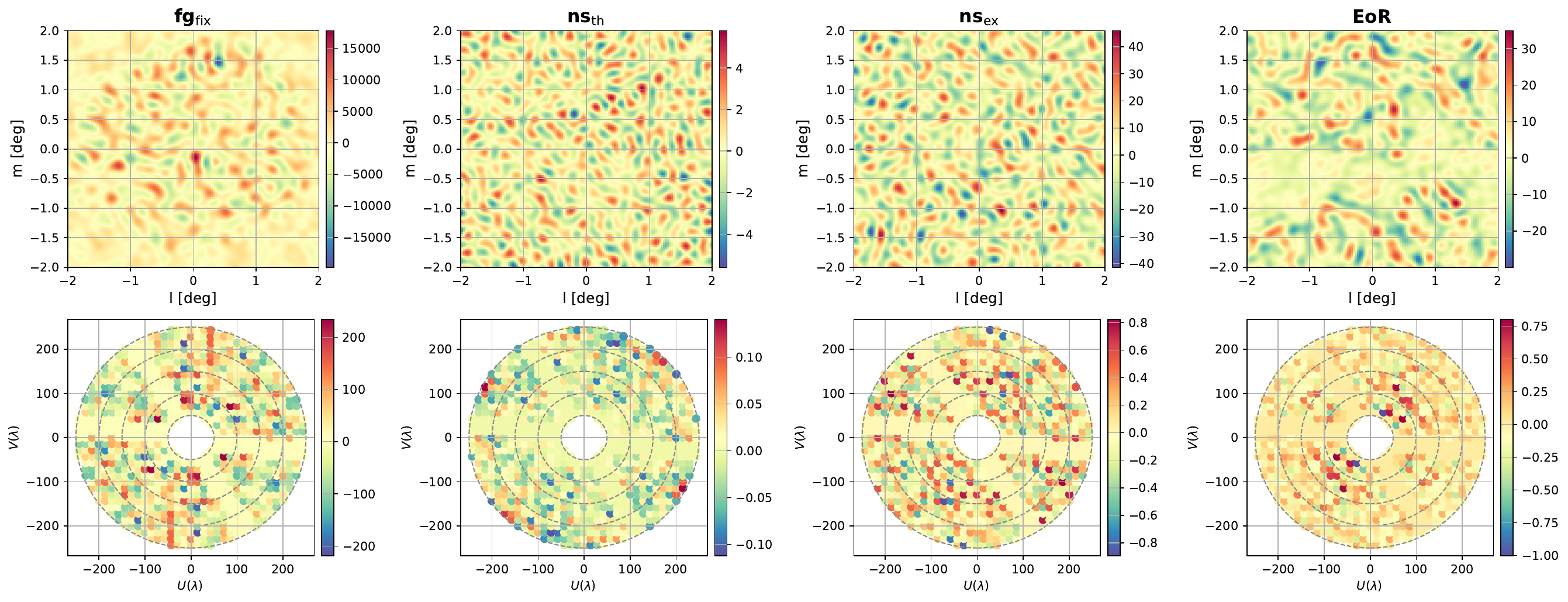}
\caption{\label{all_component_img_visi_50250}
Simulated slices of images (first line) and gridded visibilities (second line) from SKA after applying LOFAR $uv$-coverage of $\mathbf{fg}_\text{fix}$, $\mathbf{ns}_\text{th}$, $\mathbf{ns}_\text{ex}$, and $\mathbf{EoR}$ for the slices in Fig.~\ref{th_eor_component_img_visi} and Fig.~\ref{fg_ex_component_img_visi}. The units of images and gridded visibilities are both mK.}
\end{figure*}

\subsection{Mask on $uv$-coverage}
\label{Mask_on_uv-coverage}

\begin{figure*}
\centering
\includegraphics[width=0.95\textwidth]{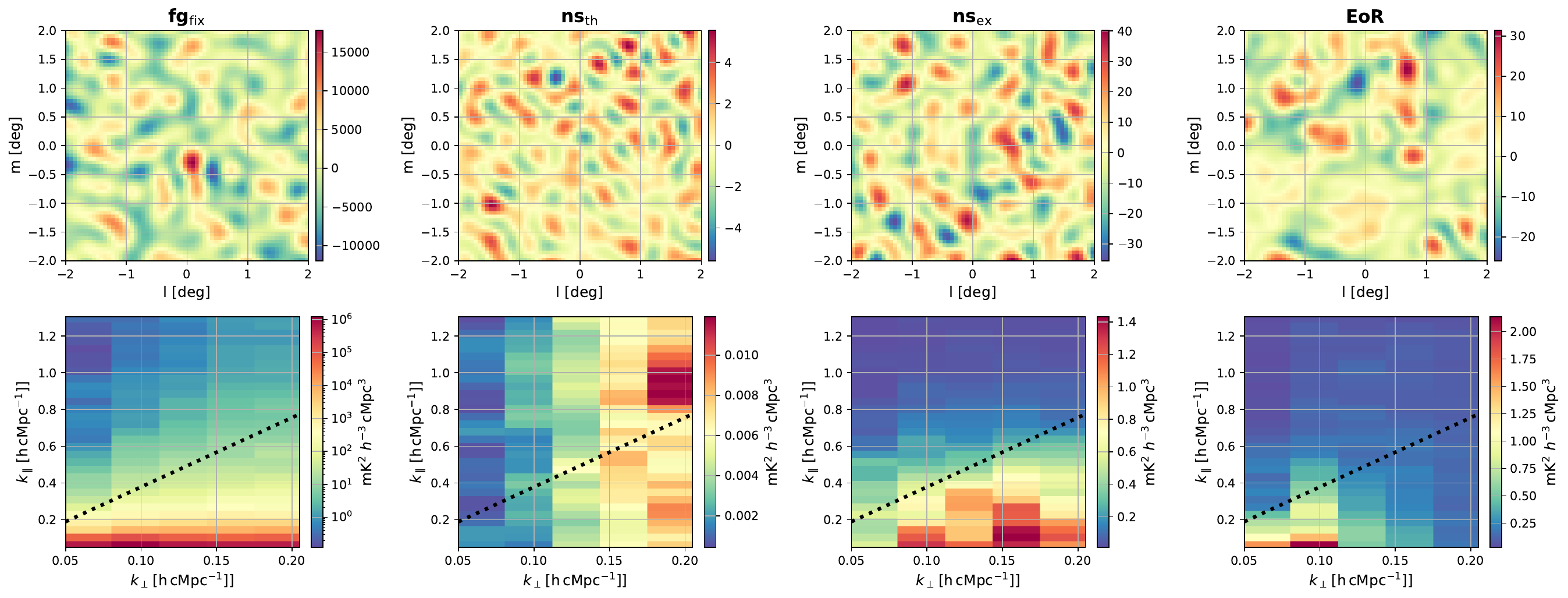}
\caption{\label{all_component_img_ps_64}
Slices of the middle $64\times64$ pixel images (first line) of the four components $\mathbf{fg}_\text{fix}$, $\mathbf{ns}_\text{th}$, $\mathbf{ns}_\text{ex}$, and $\mathbf{EoR}$ in Fig.~\ref{all_component_img_visi_50250} and their corresponding 2D power spectra (second line).
The units of images are both mK. The color bar of $\mathbf{fg}_\text{fix}$ 2D power spectrum is plotted using logarithmic normalization to better represent the dynamic range of the data. The black dotted lines are horizon lines for SKA-Low.}
\end{figure*}

As shown in Fig.~\ref{th_eor_component_img_visi} and Fig.~\ref{fg_ex_component_img_visi}, the $uv$-coverage of the foreground residuals is different from the $uv$-coverage of the other components. 
For the foreground residuals, we used the same post-flagging $uv$-coverage as in the LOFAR observations. 
In addition, there is a cross-like mask in the $uv$-coverage due to the fact that in the LOFAR observations, we flagged side-lobe residuals of CasA and CygA that appear along nearly linear lines in $uv$-space (see \cite{Munshi:2024xwv}) for a discussion of this effect).
Since there will be the same effect of high-intensity sources in the SKA observations as well, we retained this cross-like mask in our simulations of the SKA.
And for thermal noise and excess variance, the observation frequency is set to be the same as that of the foreground residuals, but the effect of a strong radio source is not taken into account. 
For the 21-cm signal, the $uv$-coverage is full between the scales corresponding to the pixel and the full image cube.
Therefore, we also need to use the cross-like mask in the observations for the $uv$-coverage of thermal noise, excess variance, and 21-cm signal which are shown in Fig.~\ref{all_component_img_visi_50250}.
The subtraction of these two strong sources will slightly reduce the signal-to-noise ratio and make the images smoother due to the loss of some of the structural information.
For the 21-cm image, in addition to considering the effect of the cross-like mask, we need to remove signals outside the $50-250 \lambda$ baseline range, so there is a significant drop in its intensity.

By comparing Fig.~\ref{th_eor_component_img_visi} and Figs.~\ref{fg_ex_component_img_visi} with Fig.~\ref{all_component_img_visi_50250}, we find that the 21-cm signal in Fig.~\ref{all_component_img_visi_50250} becomes significantly more difficult to detect.
This is because after using the $uv$-coverage of LOFAR, small-scale information in the sky map is filtered out.
The thermal noise and excess variance, on the other hand, do not change significantly in appearance.

Because of the imprint of the primary beam near the edges of the foreground residual, we selected the $128 \times 128$ part in the middle of each frequency channel for the subsequent study.
We thus get $536\times64\times128\times128$ simulated data for every component, where $536$ is the number of simulated data cubes and $64$ is the number of frequency channels.
However, this volume of about $500$ million voxels is still a challenge for subsequent deep learning in terms of GPU memory. 
We, therefore, averaged over $2\times2$ neighboring pixels in each frequency slice to end up with a final data set with the shape of $536\times64\times64\times64$.
We illustrate the slices of the components and their corresponding 2D power spectra for one of the data cubes in Fig.~\ref{all_component_img_ps_64}.
Note that, in all 2D power spectra, the black dotted lines represent the horizon lines with the flat-sky approximation based on
\begin{equation}
k_{\parallel}^{\mathrm{flat}} = k_{\perp} \frac{D_M(z) H_0 E(z)}{c(1+z)},
\end{equation}
in which $D_M(z)$ is the conversion factor from angular units to comoving distance units, $H_0$ is the Hubble constant, and $E(z) = H(z)/H_0$ is the dimensionless Hubble parameter.

\section{Signal separation via U-Net}
\label{Method}

This study employs a U-Net architecture, which is based on convolutional neural networks (CNNs) to attempt to separate the 21-cm signal from the noise, foregrounds, and excess variance components.
Due to the large memory requirements of deep learning, we utilized an NVIDIA A$100$ Tensor Core GPU with $80$ GB of memory in the DAWN compute cluster located at the Center for Information Technology of the University of Groningen \citep{2020ASPC..527..473P}.
We have $536$ data sets available for deep learning. Of these, the $512$ sets are designated for training, the $16$ sets for validation, and the remaining $8$ sets are reserved for testing.

\subsection{The U-Net architecture}
\label{The_U-Net_architecture}

The U-Net network is a widely used deep learning architecture that was initially used in tasks such as image segmentation in biomedical science \citep{ronneberger2015u}.
Due to its ability to generate an output data set with the same shape as the input while extracting essential features, U-Net has been widely utilized in the processing of sky maps \citep{Gagnon-Hartman:2021erd, Kennedy:2023zos, Bianco:2024jhe}.
In prior research using the U-Net architecture as described in \cite{Makinen:2020gvh}, we conducted various tests. 
For example, at low redshift with MeerKAT \citep{MeerKLASS:2017vgf,
2021MNRAS.501.4344L,2021MNRAS.505.3698W}, a precursor of SKA-Mid, we addressed beam effects \citep{Ni:2022kxn} as well as polarization leakage \citep{Gao:2022xdb}.

\begin{figure*}
\centering
\includegraphics[width=0.90\textwidth]{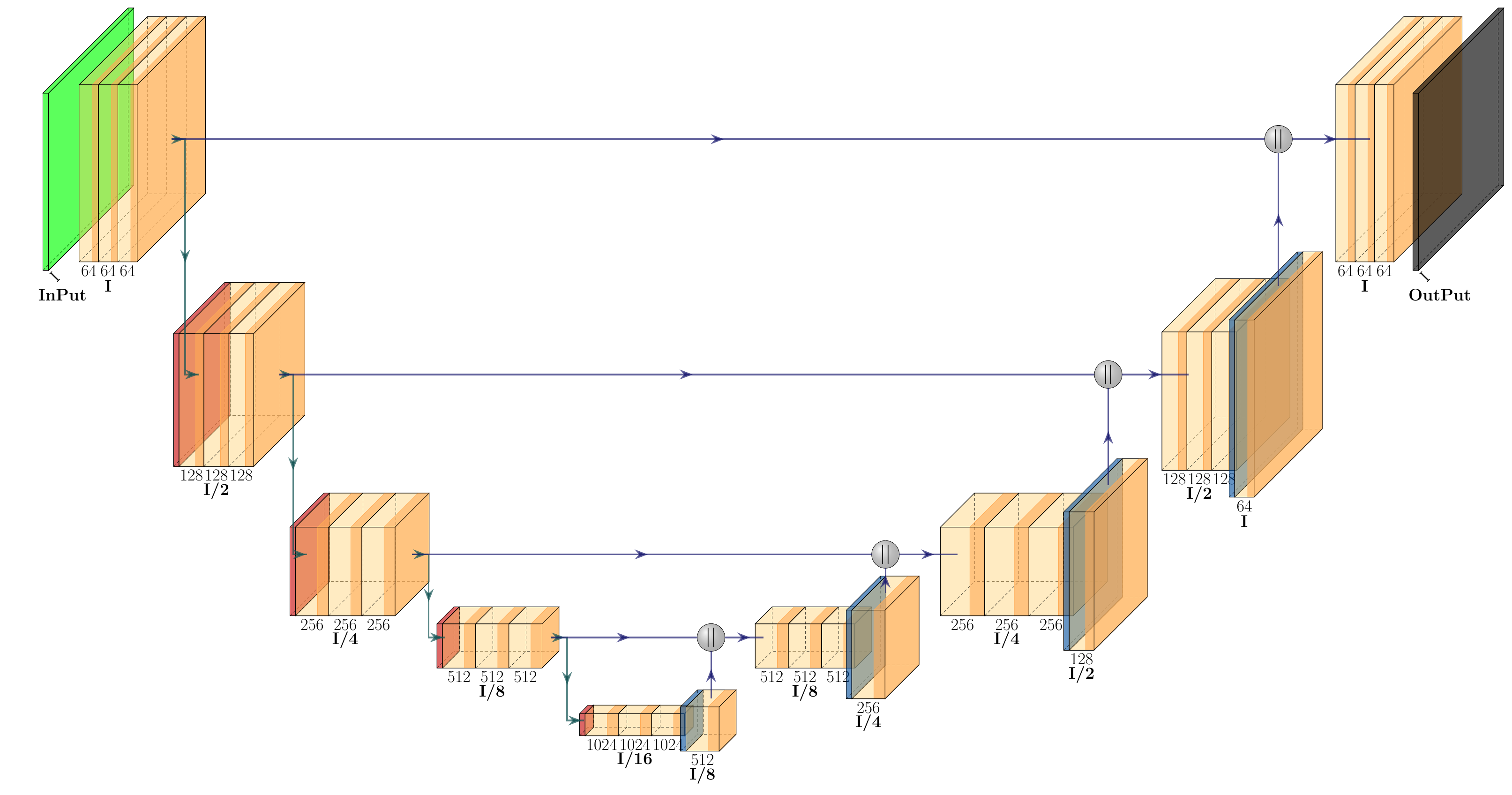}
\caption{\label{unet}Training process of CNNs with U-Net architecture. Each color represents a structure in the U-Net network, where yellow cubes represent the convolutional layers and ReLU sections, red cubes represent pooling layers in down-sampling, blue cubes represent the transposed convolutional layers, and gray spheres represent connection layers. The green and black squares at the beginning and end of this figure represent the input and output, respectively.}
\end{figure*}

The $4$-layer U-Net architecture is depicted in Fig.~\ref{unet}. 
The green square on the far left represents the input data set, while the black square on the far right denotes the output data set. 
The U-Net architecture is divided into two primary sections: the left side of the `U' is responsible for down-sampling, whereas the right side handles up-sampling.
The convolutional network in each layer of the down-sampling processing contains 3 convolutional blocks.
The first convolutional network contains $64$ convolutional kernels, and the size of each convolutional kernel is $3\times3\times3$.
To reduce information loss and reduce dimensionality more smoothly, we need to ensure that the stride is smaller than the length of the convolution kernel, so we set the stride $= 2$.
Each convolutional network is followed by a rectified linear unit (ReLU) activation.
Convolutional kernels and ReLU activations are represented within yellow boxes in Fig.~\ref{unet}. 
Following this, we employ a maximal set operation (red box), which can also be referred to as a pooling layer. 
This down-sampling process condenses the structural information of the input data cube into a reduced set of features.
Since we set a growth factor of 2, each pooling operation results in a halving of the spatial dimension, and the number of channels increases by a factor of 2 of the previous layer.
The parameters related to the down-sampling processing are listed in Table \ref{Hyper_unet}.
For the up-sampling process, the blue part represents the transposed convolution and the gray sphere denotes the connected layer.
The transposed convolution operation is capable of expanding the spatial dimensions of a feature map from a lower resolution to a higher resolution. 
The connected layer links the corresponding layers involved in both down-sampling and up-sampling processes, thereby preventing information loss that can occur with increasing network depth and aiding in the retention of small-scale image structures.
Throughout the entire learning process, we employ the {\tt AdamW} optimizer \citep{Loshchilov:2017bsp} to achieve a stepwise reduction in the learning rate.

\subsection{Loss function}

The loss function in deep learning quantifies the disparity between the predicted values generated by a network and the actual true values.
The standard Mean Square Error (MSE) loss function is the most frequently utilized; however, it exhibits substantial instability when processing images.
Consequently, we employ the more stable Log-Cosh loss function, which enhances the management of outlier points.
The Log-Cosh loss function is defined as
\begin{equation}
\mathcal{L}=\sum\limits_{i} \log\cosh(p_i - t_i),
\label{loss_fun}
\end{equation}
where $p_i$ represents the predicted outcome and $t_i$ corresponds to the actual signal of the $i$-th voxel, respectively.
We also considered calculating the loss in $uv$-coverage instead of image space.
However, since there is no obvious relationship between neighboring pixels in $uv$-coverage, the results obtained are significantly worse than operating in image space.

\subsection{Hyperparameter selection}
\label{Hyperparameter_selection}
Hyperparameters determine key aspects of network architecture, learning process, optimization methods, and regularization strategies.
We list the hyperparameters used by U-Net in Table \ref{Hyper_unet}. 
We have already described the setting of the hyperparameters related to the convolutional and pooling layers of the deep learning network used in Section~\ref{The_U-Net_architecture}.
We configure $n_{\rm down}$ to $4$ to allow the network to access all data.
Since the 21-cm signal is extremely weak, our primary focus is on utilizing the most fine-tuned neural network that the GPU memory capacity can support, rather than optimizing for speed. 
Setting $n_{\rm block}$ to 3 ensures that we get more information about the features and prevents loss of information.
To facilitate the use of complex neural network structures, we maintain a relatively modest common batch size of $48$.
We utilize $64$ initial convolution filters per block to thoroughly capture detailed features.
We use a smaller learning rate ($\eta = 10^{-5}$) to mitigate overfitting, complemented by the {\tt AdamW} optimizer \citep{Loshchilov:2017bsp}, which adjusts the learning rate during training to further prevent overfitting.
The weight decay $\omega$ in the optimizer is set to $10^{-5}$ to ensure that the later training process has a small enough learning rate in order to prevent overfitting.
The batch normalization momentum ($\beta_{\rm mom}$) is set to $0.02$, ensuring that the current batch statistics are balanced against historical estimates to improve the neural network's stability.

\begin{table}
\begin{center}
\caption{\label{Hyper_unet}Description of the hyperparameters in the U-Net architecture design.}
	\begin{tabular}{llcc}
		\hline
		Hyperparameter                                             & Value\\
		\hline
        convolution width~(number of convolutions in each layer)   & 3         \\
        kernel size~(size of the convolution kernel)               & $3\times3\times3$           \\
        growth factor~(Growth rate of channels in each layer)      & 2           \\
        stride~(step size of each move of the convolution kernel)  & 2         \\
		$n_{\rm block}$~(number of convolutions for each block)    & 3           \\
		$n_{\rm down}$~(number of down-convolutions)               & 4           \\
		batch size~(number of samples per gradient descent step)   & 48          \\
		$n_{\rm filter}$~(initial number of convolution filters)   & 64          \\
        $\eta$~(learning rate for optimizer)                       & $10^{-5}$   \\
		$\Omega$~(optimizer for training)                          & {\tt AdamW} \\
        $\omega$~(weight decay for optimizer)                      &$10^{-5}$    \\
		$\beta_{\rm mom}$~(batch normalization momentum)           & $0.02$      \\                               
		\hline
	\end{tabular}
\end{center}
\end{table}

\section{Results and discussion}
\label{Results_and_discussion}

We tested the results of 3D U-Net on LOFAR mock data and show them in Appendix~\ref{appendixB} and Appendix~\ref{appendixC}.
However, due to the high noise level of LOFAR, we were not able to extract the 21-cm signal efficiently.
In this section, we examine the ability of 3D U-Net to extract the 21-cm signal based on different components and observation times of the mock SKA data.
Furthermore, we conduct a robustness analysis to assess to what level the 3D U-Net can recover the 21-cm signal in the foreground-wedge region.

\subsection{Signal extraction from $\mathbf{ns}_\text{th}$ + $\mathbf{EoR}$}

\begin{figure}
\centering
\includegraphics[width=0.45\textwidth]{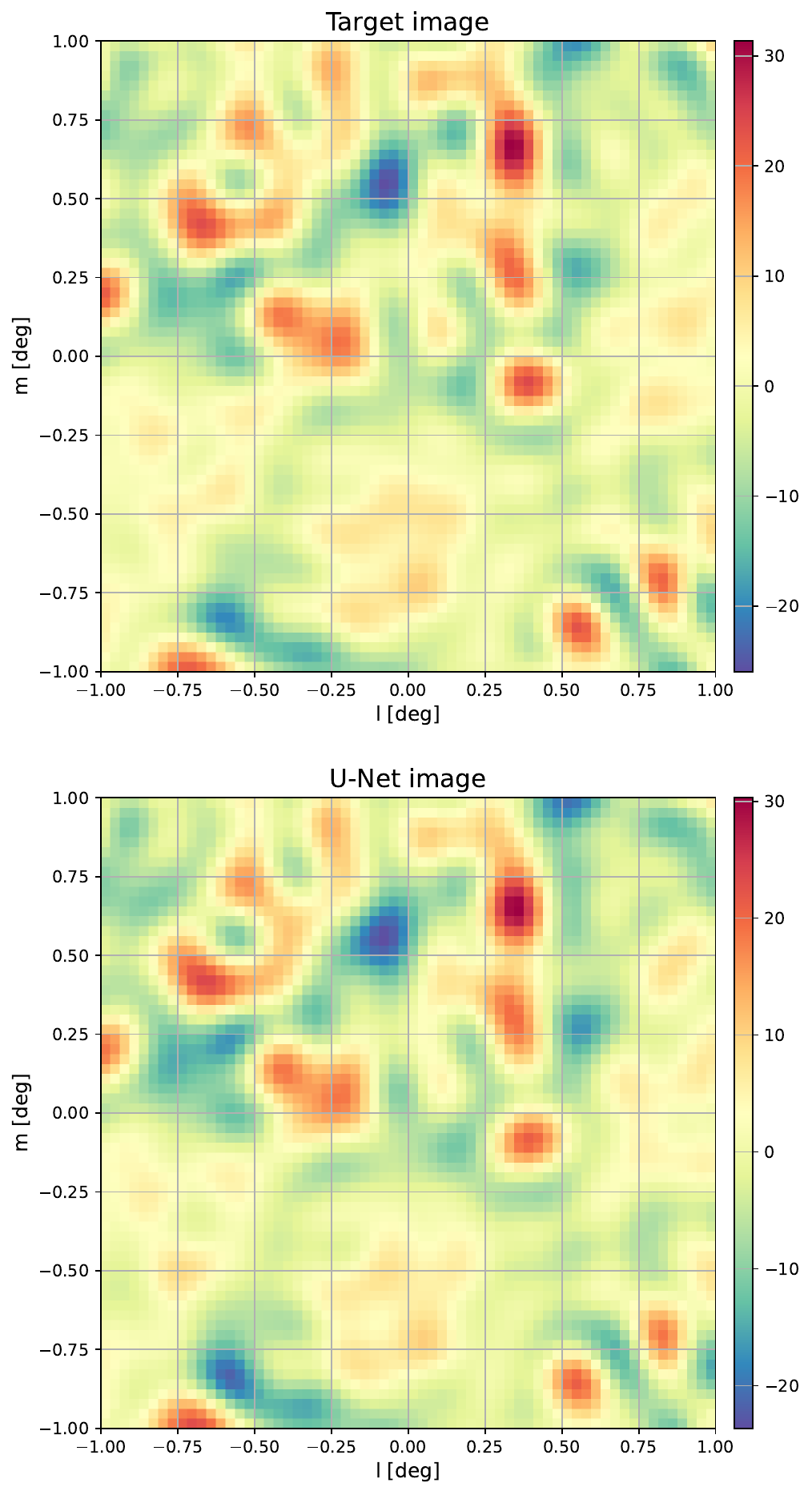}
\caption{\label{nseor_image}Target $\mathbf{EoR}$ image and predictive images given by U-Net when considering only the effects of $\mathbf{ns}_\text{th}$. These images are in units of mK.}
\end{figure}

\begin{figure}
\centering
\includegraphics[width=0.49\textwidth]{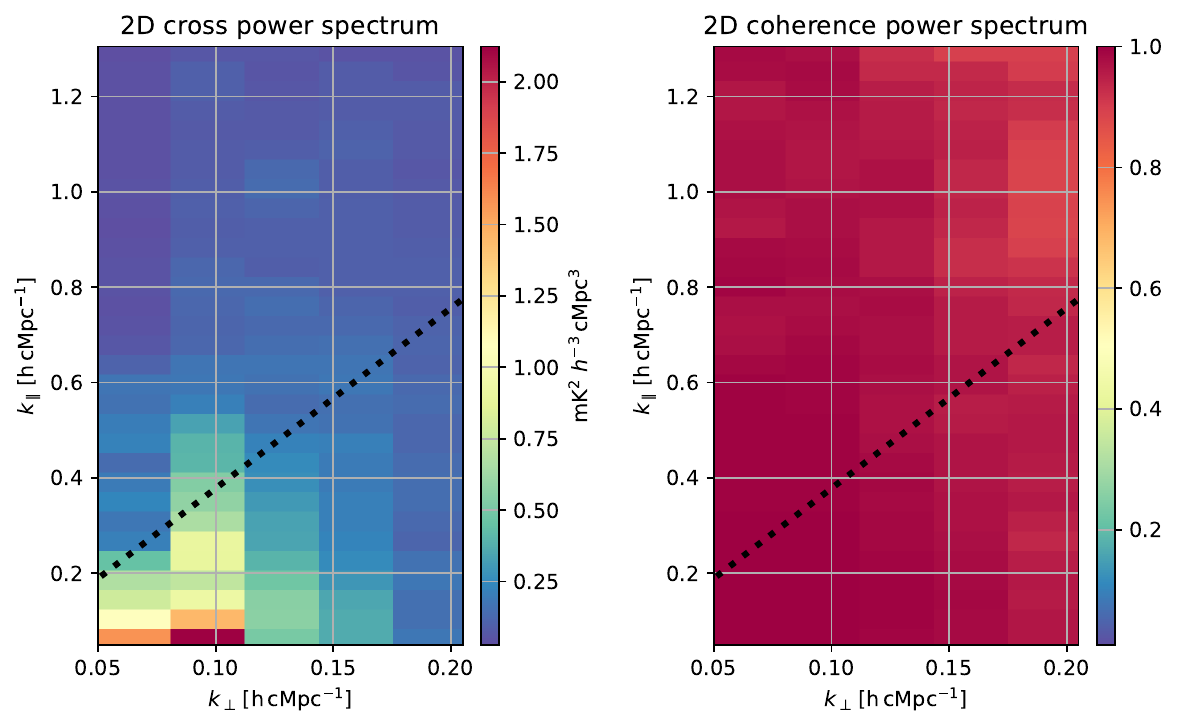}
\caption{\label{nseor_spectra2D}2D cross power spectrum and 2D coherence power spectrum between the target $\mathbf{EoR}$ and U-Net predictive image when considering only the effects of $\mathbf{ns}_\text{th}$. The black dotted lines are horizon lines for SKA-Low.}
\end{figure}

Based on 1752 hours of SKA-Low observations, both $\mathbf{fg}_\text{fix}$ and $\mathbf{ns}_\text{ex}$ far exceed the 21-cm signal, while $\mathbf{ns}_\text{th}$ is an order of magnitude below the 21-cm signal on most baselines simulated in this work.
As a first step, we evaluate a simplified scenario, focusing solely on the 21-cm signal and $\mathbf{ns}_\text{th}$, while disregarding $\mathbf{fg}_\text{fix}$ and $\mathbf{ns}_\text{ex}$.

We find that the 3D U-Net can easily extract the required information, since the thermal noise level is small relative to the 21-cm signal, allowing U-Net to discern finer structures. 
Consequently, we require more training epochs to achieve optimal results. 
After 7000 epochs, the loss-function value of the U-Net stabilizes. 
Although further training does not result in overfitting, the network is becoming slightly unstable.

It is important to note that varying data combinations, each with unique structures and errors, require different numbers of training epochs to maintain stability and prevent overfitting, as will be demonstrated next. 
In order to quantitatively evaluate the performance of the 3D U-Net in extracting the 21-cm signal, we define the 2D corss power spectrum as
\begin{equation}
C_{1,2}^{\rm cross}(k_\perp, k_\parallel) \equiv \left\langle \tilde{T}_1^*(\mathbf{k}) \tilde{T}_2(\mathbf{k}) \right\rangle,
\end{equation}
and the 2D coherence power spectrum as
\begin{equation}
C_{1,2}^{\rm coherence}(k_\perp, k_\parallel) \equiv \frac{\left\langle \tilde{T}_1^*(\mathbf{k}) \tilde{T}_2(\mathbf{k}) \right\rangle^2}{\left\langle \left| \tilde{T}_1(\mathbf{k}) \right|^2 \right\rangle \left\langle \left| \tilde{T}_2(\mathbf{k}) \right|^2 \right\rangle},
\end{equation}
in which the indices 1 and 2 represent the target image and the U-Net result, respectively.
The target 21-cm signal image and the U-Net output are illustrated in Fig.~\ref{nseor_image} for a typical case. 
Although there are some small differences, U-Net recovers most structures successfully. 
This conclusion is supported by the 2D cross power spectrum and 2D coherence power spectrum between the target $\mathbf{EoR}$ and U-Net images, as shown in Fig.~\ref{nseor_spectra2D}.
We see that the coherence is very close to unity and only for a higher k-mode decreases a little due to the thermal noise.

\subsubsection{Robustness analysis}

\begin{figure}
\centering
\includegraphics[width=0.45\textwidth]{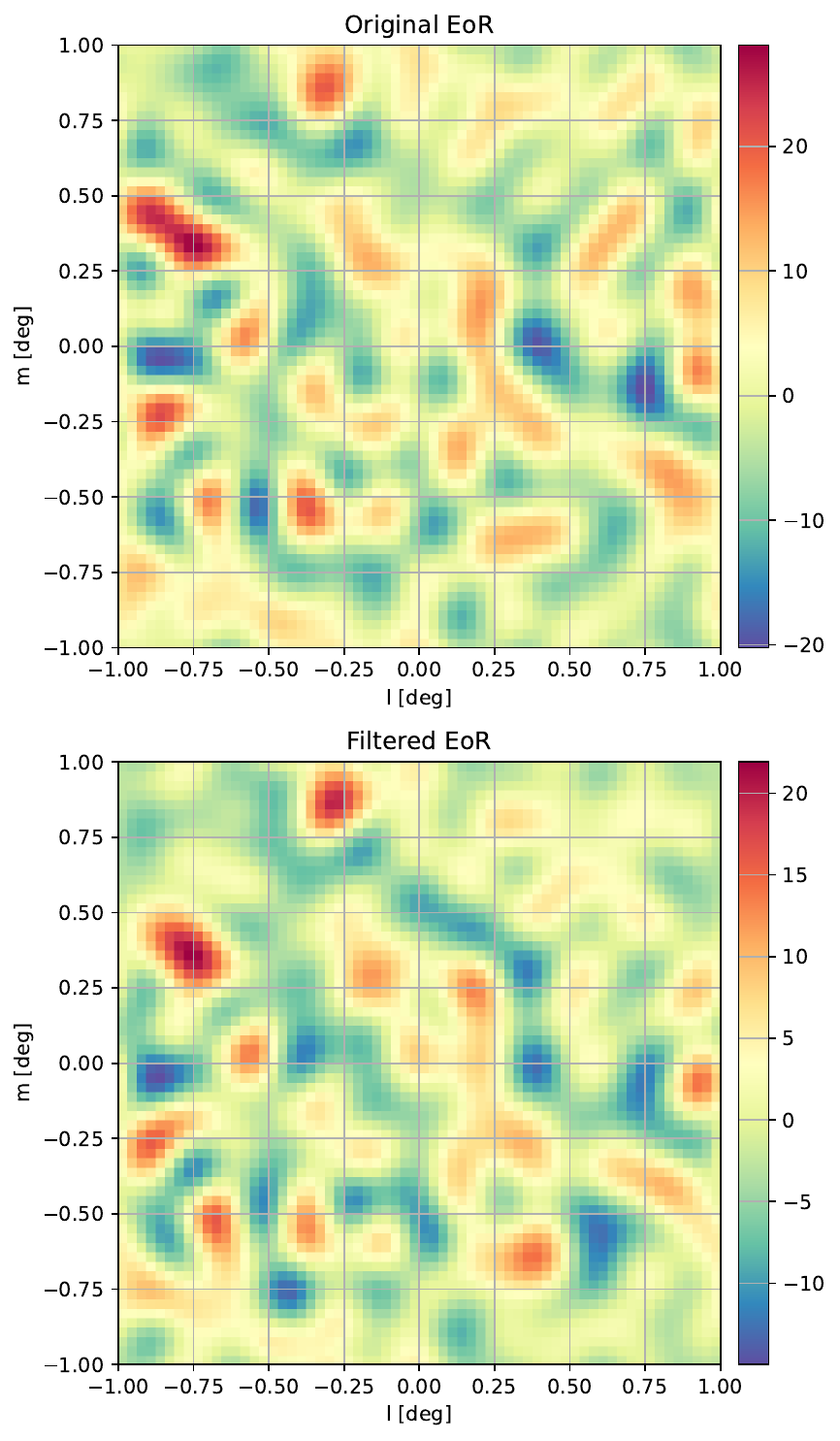}
\caption{\label{filtered_image}Original $\mathbf{EoR_{ori}}$ image and regenerated $\mathbf{EoR_{rev}}$ image by removing the $30^\circ$ filter wedge in the 2D power spectrum. These images are in units of mK.}
\end{figure}

\begin{figure*}
\centering
\includegraphics[width=0.9\textwidth]{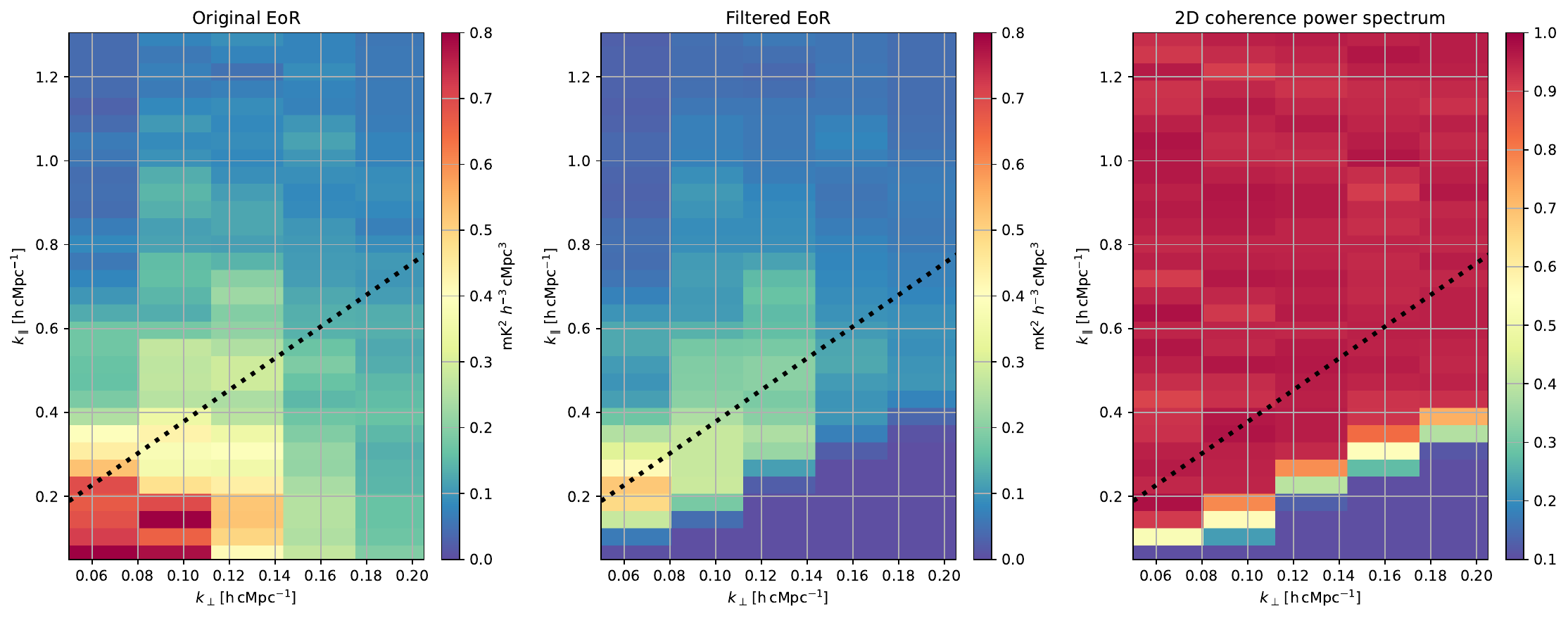}
\caption{\label{filtered_ps}Original $\mathbf{EoR_{ori}}$ 2D power spectrum, $\mathbf{EoR_{rev}}$ 2D power spectrum after removing the $30^\circ$ filter wedge, and their 2D coherence power spectrum between them. We find that these two spectra are different in scale before and after filtering. Due to windowing effects based on the finite frequency bandwidth of the observations, it leads to some leakage from scales below the wedge to above the wedge after, and vice versa. Hence, removing modes below the wedge leads to a minor change in both power and coherence above the wedge as well. The black dotted lines are horizon lines for SKA-Low.}
\end{figure*}

\begin{figure}
\centering
\includegraphics[height =0.45\textwidth]{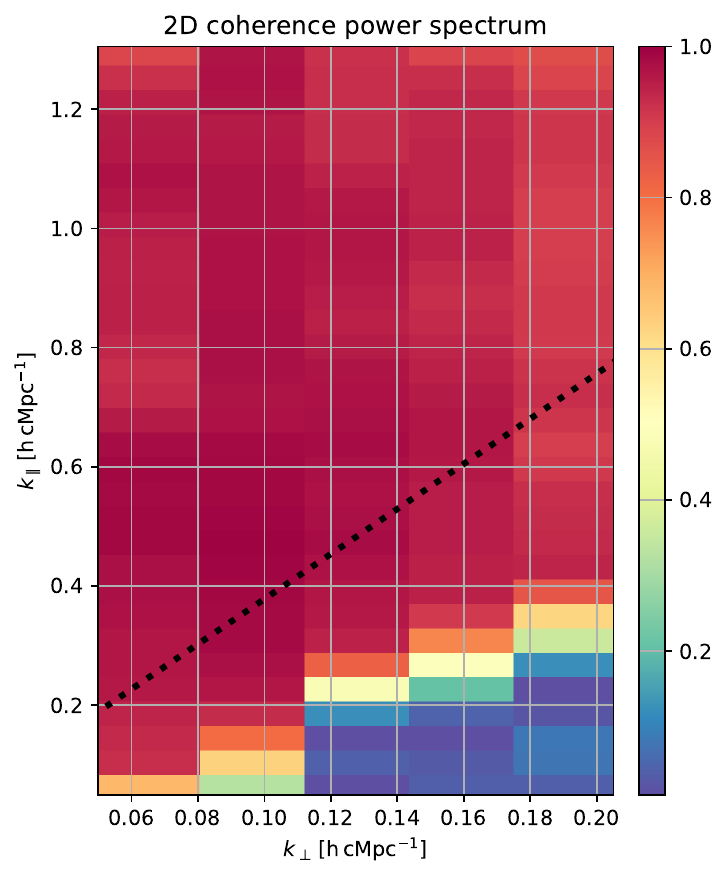}
\caption{\label{coherence_ps_rob}2D coherence power spectrum of U-Net predicted $\mathbf{EoR_{rev}}$ image and target $\mathbf{EoR_{ori}}$ image. The black dotted lines are horizon lines for SKA-Low.}
\end{figure}

In the following subsections, we will assess the impact of the foregrounds ($\mathbf{fg}_\text{fix}$) and the excess variance ($\mathbf{ns}_\text{ex}$) on the recovery of the 21-cm signal using U-Net, showing that recovery becomes more difficult.  However, U-Nets are also capable of predicting signals in regions where they are not measured (e.g., inside the wedge) using nearby information  (e.g., outside the wedge) \citep{Gagnon-Hartman:2021erd, Kennedy:2023zos}. 
It is important to note that they only predicted the shape and location of ionized regions, whereas in this work, we decided to fully predict all the information in the images.
We therefore first need to assess whether our U-Net predicts or genuinely recovers the 21-cm signal inside the wedge region.

The power of the foreground is concentrated within a $30^\circ$ wedge region in the 2D power spectrum, and we wish to simulate the scenario of subtracting the foreground-dominated part of the power spectrum.
To confirm this, we used a filter wedge with a $30^\circ$ angle. 
For $\mathbf{EoR}$, the part of the power spectrum within the filter wedge is excluded, resulting in a revised $\mathbf{EoR}$ sky map.
To distinguish them, we call them $\mathbf{EoR_{ori}}$ and $\mathbf{EoR_{rev}}$ respectively 
We present both the $\mathbf{EoR_{ori}}$ image and the $\mathbf{EoR_{rev}}$ image obtained after removing the filter wedge from the 2D power spectrum in Fig.~\ref{filtered_image}. 
Furthermore, Fig.~\ref{filtered_ps} presents their respective 2D power spectra along with their 2D coherence power spectrum.
As demonstrated in Fig.~\ref{filtered_image}, the peak intensity of the $\mathbf{EoR_{rev}}$ decreases considerably with the removal of the wedge, but they remain very similar in their overall structure. 

We utilized the original $\mathbf{ns}_\text{th}$ image combined with the $\mathbf{EoR_{rev}}$ image as the input, with the $\mathbf{EoR_{ori}}$ image serving as the target, and fed them into the 3D U-Net model.
After 7000 epochs of training, the coherence 2D power spectrum between the U-Net predicted skymap and the target skymap was obtained as shown in Fig.~\ref{coherence_ps_rob}.
Continuing the training further would lead to instability in the neural network.
By analyzing the coherence power spectra in Figs.~\ref{filtered_ps} and \ref{coherence_ps_rob}, we conclude that the neural network does not fully predict the filtered-out part, i.e., in the wedge, of the power spectrum when recovering the full images, which is different from the results of \cite{Gagnon-Hartman:2021erd, Kennedy:2023zos}. 
This is due to the fact that U-Net is sufficient for an image segmentation task, whereas for the full recovery of $\mathbf{EoR_{ori}}$ image, U-Net is unable to learn the structure from the image due to the lack of the corresponding signal.
Consequently, we conclude that any recovered 21-cm signal below the wedge in the presence of strong foreground and excess variance, as presented in the following sub-sections, is not due to a prediction from signal above the wedge, but a genuine signal recovery.

\subsection{Signal extraction from $\mathbf{fg}_\text{fix}$ + $\mathbf{ns}_\text{th}$ + $\mathbf{EoR}$}
\label{fgnsEoR}

Here we examine the influence of the fixed foreground residual $\mathbf{fg}_\text{fix}$ and thermal noise $\mathbf{ns}_\text{th}$, on the signal extraction. 
Given that the foreground remains fully coherent during the observation period, it is primarily excess variance and thermal noise that impact the observations during training, and the foreground residuals are based on the foreground residuals (after sky-model subtraction) from LOFAR observations. 
Therefore, we treat the foreground residual as fixed.
In order to further evaluate the U-Net's capability to extract the $\mathbf{EoR}$ signal, for now, we assume ideal observations without any systematic effects such as the excess variance due to mode mixing.
It should be noted that in this training process, the input consists of $\mathbf{fg}_\text{fix}$ + $\mathbf{ns}_\text{th}$ + $\mathbf{EoR}$. 

\begin{figure}
\centering
\includegraphics[width=0.49\textwidth]{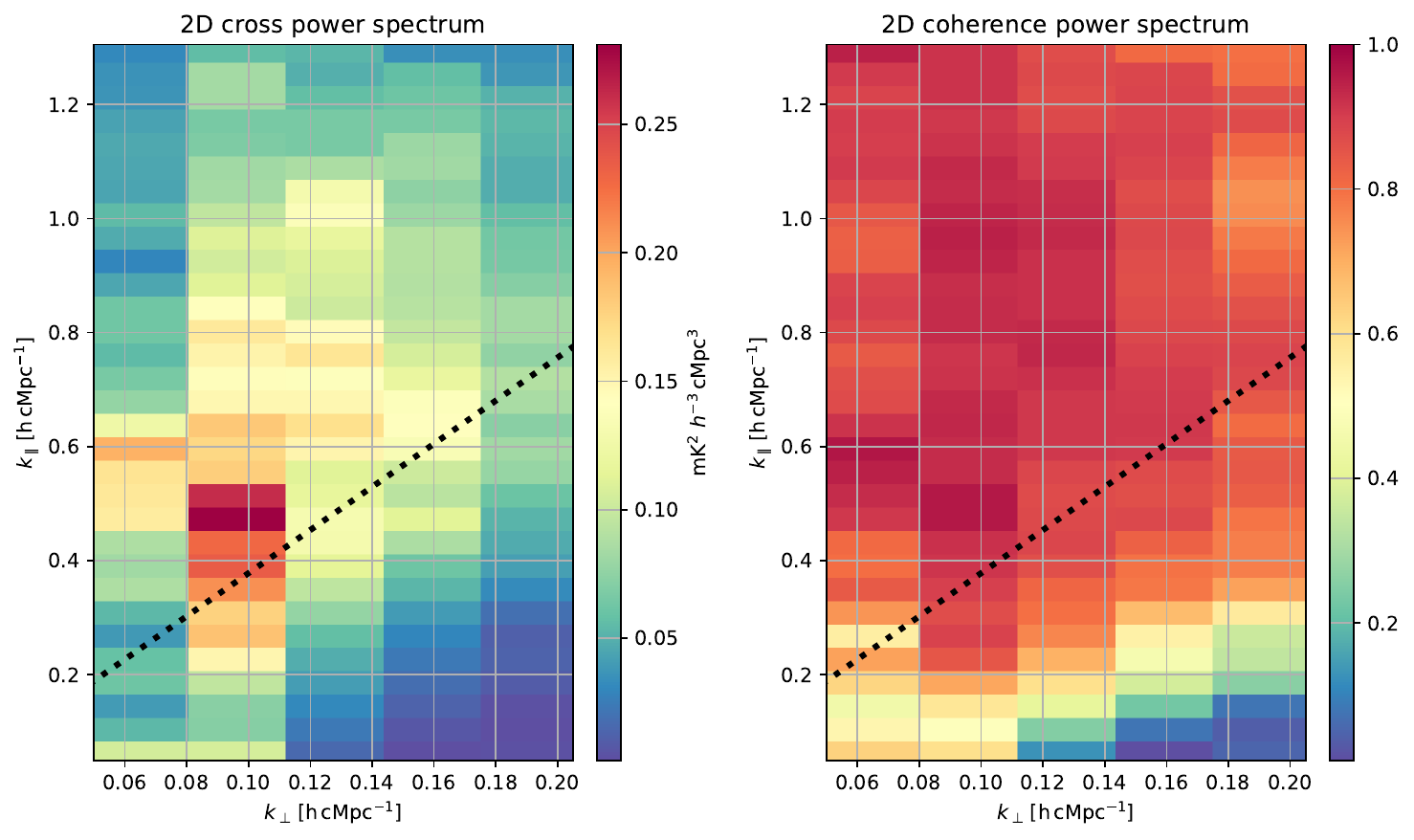}
\caption{\label{fix_ns_eor_ps}2D cross power spectrum and 2D coherence power spectrum between the target $\mathbf{EoR}$ and U-Net predictive image when considering the effects of $\mathbf{fg}_\text{fix}$ and $\mathbf{ns}_\text{th}$. The black dotted lines are horizon lines for SKA-Low.}
\end{figure}

Fig.~\ref{fix_ns_eor_ps} shows the 2D cross power spectrum and 2D coherence power spectrum between the target $\mathbf{EoR}$ image and the U-Net predicted image from 1500-epoch training after including the fixed foreground model.
Further training causes the loss-function to become significantly unstable.
Comparing Fig.~\ref{nseor_spectra2D} and Fig.~\ref{fix_ns_eor_ps} above the horizon, we see that both give excellent 21-cm signal recovery over much of the probed spatial scales. 
It shows that for small-scale structures, the fixed foreground residuals do not affect the training results.
However, for a small section below the horizon, there is a more pronounced inconsistency in its lower right corner in Fig.~\ref{fix_ns_eor_ps}, which is consistent with the signal extraction results from other methods.
Examining the power spectrum of the foreground residual illustrated in Fig.~\ref{all_component_img_ps_64} alongside the 2D coherence power spectrum in Fig.~\ref{fix_ns_eor_ps} reveals that U-Net is unable to accurately reproduce the power spectrum in areas where the intensity of the foreground residuals is high, despite the fixation of the foreground residuals we included in the analysis.
 To solve this problem, we will examine a combination of using GPR and U-Nets in a future publication since the GPR in general is quite effective in removing those modes from the data, and this seems to be necessary for the U-net to perform optimally.

\subsection{Signal extraction from $\mathbf{ns}_\text{ex}$ + $\mathbf{ns}_\text{th}$ + $\mathbf{EoR}$ with different observation time}
\label{differtime}

\begin{figure*}
\centering
\includegraphics[width=0.90\textwidth]{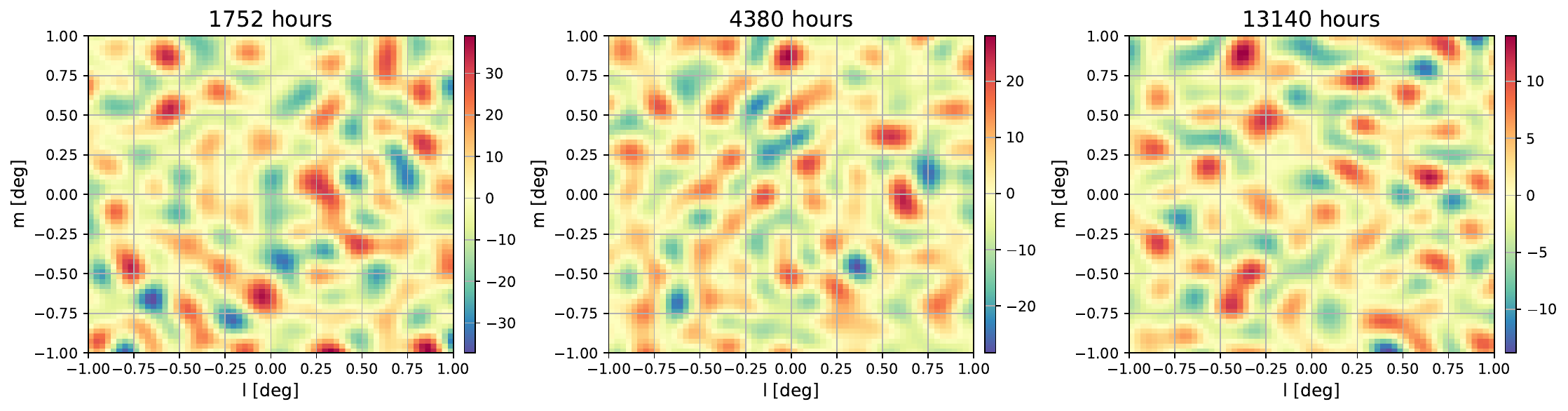}
\caption{\label{ns_image}Images of the $\mathbf{ns}_\text{all}$ at observation times 1752 hours, 4380 hours, and 13140 hours, respectively. These images are in units of mK.}
\end{figure*}

Here we have considered the effect of the excess variance ($\mathbf{ns}_\text{ex}$) in addition to the thermal noise ($\mathbf{ns}_\text{th}$).
Since the effect of $\mathbf{fg}_\text{fix}$ on the extraction of the 21-cm signal is mainly in the small $k_{\parallel}$ region and can probably be removed using GPR, we decided to ignore the effect of $\mathbf{fg}_\text{fix}$ to prevent overstressing U-Net processing in this part.
Note that the $\mathbf{ns}_\text{ex}$ we use here is a Gaussian random field, but in real observations, the $\mathbf{ns}_\text{ex}$ could have some correlated structures. 
Since Gaussian $\mathbf{ns}_\text{ex}$ makes the distribution of the intensities uncorrelated in phase space, we believe that in real observations better results could be obtained.

To determine at what noise level U-Net remains effective, we varied the duration of the observation. 
Initially, we examined a 1752-hour observation for consistency with the previous analysis, followed by observations of 4380 hours and an extreme case of 13140 hours.
For convenience, we refer to the sum of $\mathbf{ns}_\text{ex}$ and $\mathbf{ns}_\text{th}$ as $\mathbf{ns}_\text{all}$.
As the observation time increases, the excess variance decreases. 
For example, the intensity at 4380 hours of $\mathbf{ns}_\text{all}$ is about half that of 1752 hours of $\mathbf{ns}_\text{all}$, and at 13140 hours, the intensity is approximately $70\%$ of what it is at 4380 hours of $\mathbf{ns}_\text{all}$.
In Fig.~\ref{ns_image}, we illustrate some of the $\mathbf{ns}_\text{all}$ images obtained with different observation times.
Observations over 1752 hours show a peak amplitude of approximately $40$mK, while those over 4380 hours exhibit an amplitude close to $25$mK. For the 13140-hour observations, the maximum amplitude is about $15$mK, in all cases dominated by the excess variance component.
We note that the variation in the level of thermal and excess noise influences the required number of epochs to achieve stability in U-Net training.
For the observation of 1752 hours, increasing the number of epochs, however, can result in overfitting. 
This occurs because weaker structures are overwhelmed by Gaussian random field errors $\mathbf{ns}_\text{all}$, and when U-Net tries to capture the structure after noise saturation, it results in larger errors.
Observations lasting 1752 hours require 1400 epochs for U-Net, while 4380 hours of observations require 2000 epochs, and a total of 13140 hours of observations require 2500 epochs.

\begin{figure*}
\centering
\begin{tikzpicture}
\scope[nodes={inner sep=0,outer sep=0}]
\node[anchor=south east] (a)
  {\includegraphics[width=1\textwidth]{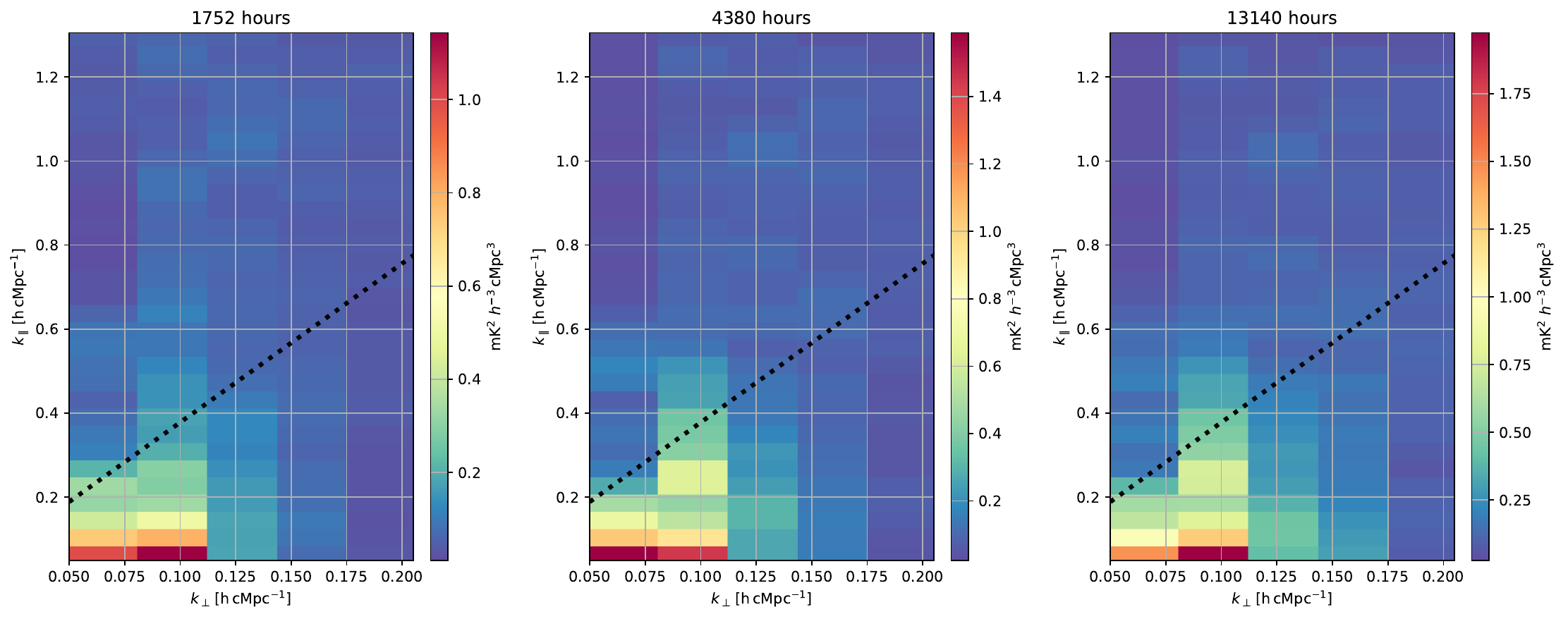}};
\node[below=0mm of a] (b)
  {\includegraphics[width=1\textwidth]{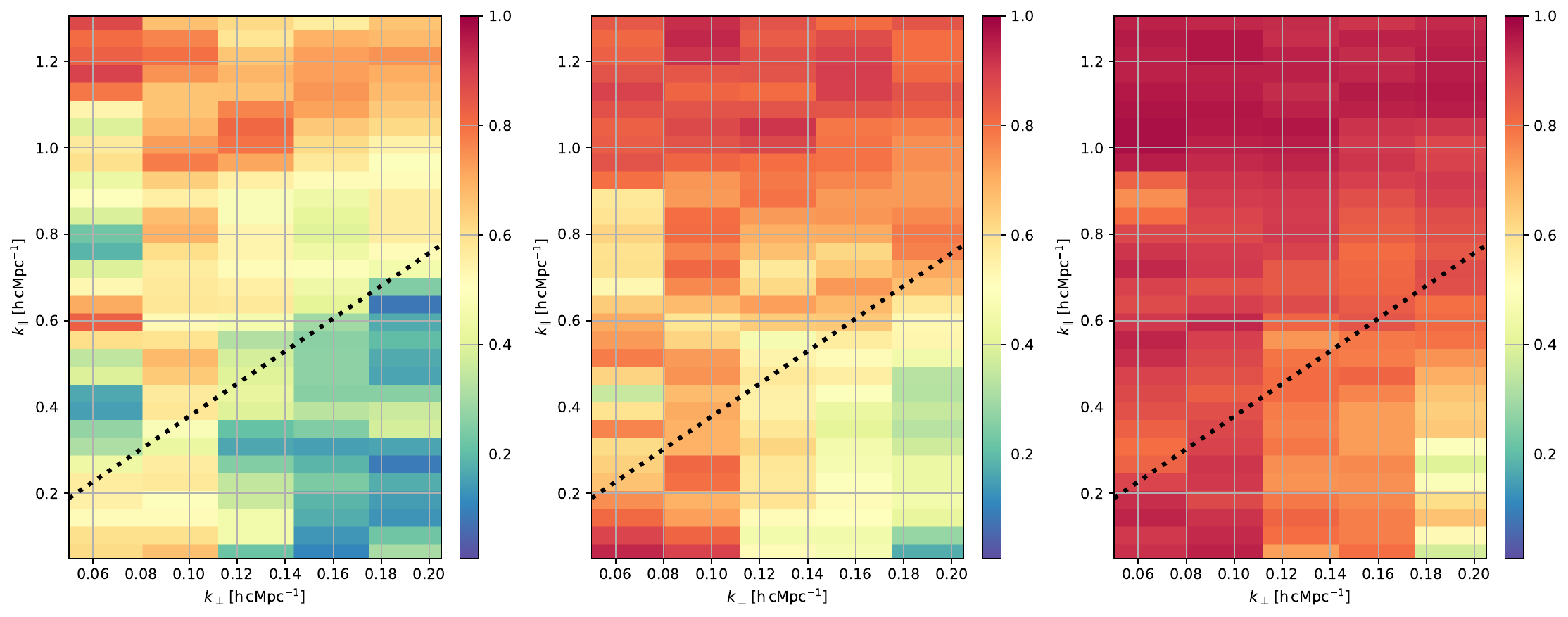}};
\endscope
\end{tikzpicture}
\caption{\label{diff_obs_cross}
The 2D cross power spectra (first row) and their corresponding 2D coherence power spectra (second row) of predicted and target images derived from U-Net processing of simulated data ($\mathbf{ns}_\text{ex}$ + $\mathbf{ns}_\text{th}$ + $\mathbf{EoR}$), which were observed for durations of 1752 hours, 4380 hours, and 13140 hours. The black dotted lines are horizon lines for SKA-Low.}
\end{figure*}

In Fig.~\ref{diff_obs_cross}, we show the 2D cross power spectra and 2D coherence power spectra of the predicted and target $\mathbf{EoR}$ images obtained via U-Net based on images at different observation times, respectively.
And the mean values of the coherence power spectra based on observations over periods of 1752 hours, 4380 hours and 13140 hours are 0.49, 0.68, and 0.85, respectively.
Our results indicate that across different observations, better outcomes are typically observed at higher $k_{\parallel}$ and lower $k_{\perp}$, as expected, since the impacts of the thermal noise and excess variance are smaller in those regions. 
Notably, for 1752 hours of observation, there is a marked alteration in the 2D coherence power spectrum when $k_{\perp}$ equals $0.113$. 
This shift may be attributed to the rapidly increasing dominance of the $\mathbf{ns}_\text{all}$ intensity over the $\mathbf{EoR}$ signal, leading to the masking of fine details in the 21-cm signal by Gaussian errors, restricting the U-Net to mainly capture larger-scale information. 
And $\mathbf{ns}_\text{ex}$ has no significant correlation in the frequency direction, which makes U-Net unable to extract the signal effectively.
In the case of 4380 hours of observation, the delimitation on $k_{\perp}$ becomes less distinct; however, the 2D coherence power spectrum still exhibits inconsistencies below the horizon line (black dotted line), similar to those in real observations. 
Meanwhile, improvements are noted for signals above the horizon line. 
Finally, after 13140 hours of observations, the intensity of $\mathbf{ns}_\text{all}$ is roughly half that of the $\mathbf{EoR}$ signal, allowing for a reliable $\mathbf{EoR}$ 2D cross power spectrum even below the horizon line.
We note, however, that obtaining 13140 hours of observations with SKA-Low on deep fields is not a likely scenario and a more effective way to recover the signal with shorter integration times would be to more effectively reduce excess variance, which we assumed here to be the same between LOFAR and SKA-Low, be incoherent, and scale down with thermal noise in the same way. 
It is very likely that SKA-Low will have less excess variance due to its better beam control and better instantaneous $uv$-coverage, leading to lower gain errors.

\subsection{Signal extraction from $\mathbf{fg}_\text{fix}$ + $\mathbf{ns}_\text{ex}$ + $\mathbf{ns}_\text{th}$ + $\mathbf{EoR}$ with different observation time}
\label{differtimefg}

\begin{figure*}
\centering
\begin{tikzpicture}
\scope[nodes={inner sep=0,outer sep=0}]
\node[anchor=south east] (a)
  {\includegraphics[width=1\textwidth]{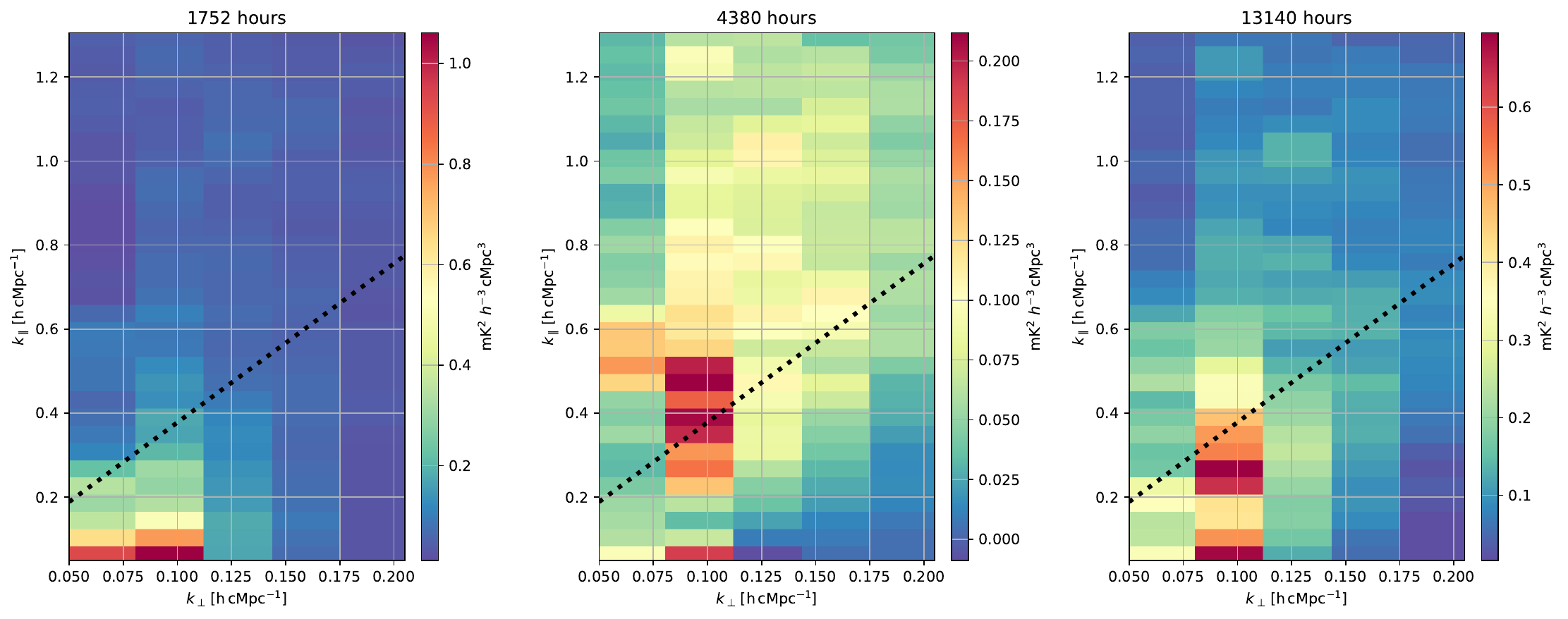}};
\node[below=0mm of a] (b)
  {\includegraphics[width=1\textwidth]{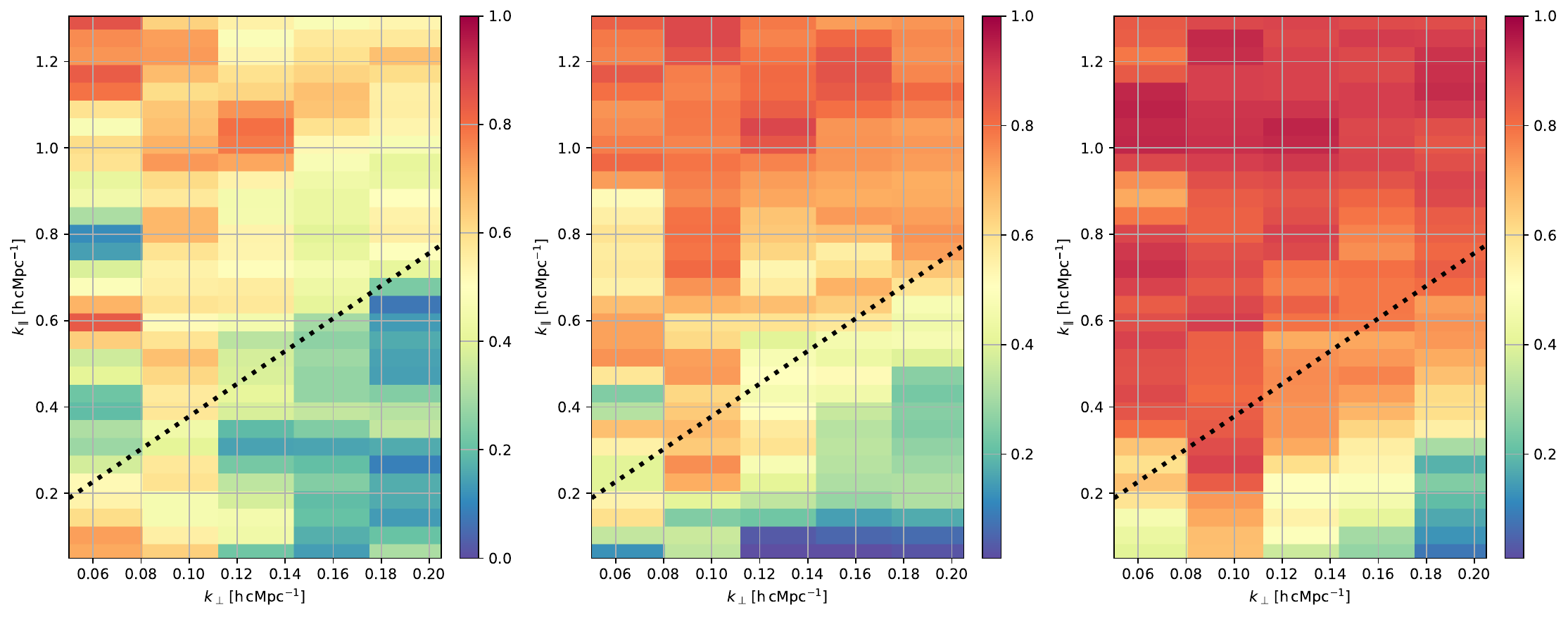}};
\endscope
\end{tikzpicture}
\caption{\label{diff_obs_cross_fg}
The 2D cross power spectra (first row) and their corresponding 2D coherence power spectra (second row) of predicted and target images derived from U-Net processing of simulated data ($\mathbf{fg}_\text{fix}$ + $\mathbf{ns}_\text{ex}$ + $\mathbf{ns}_\text{th}$ + $\mathbf{EoR}$), which were observed for durations of 1752 hours, 4380 hours, and 13140 hours. The black dotted lines are horizon lines for SKA-Low.}
\end{figure*}

Finally, we tested the most realistic scenario, i.e., considering the effects of $\mathbf{fg}_\text{fix}$, $\mathbf{ns}_\text{ex}$, and $\mathbf{ns}_\text{th}$ on $\mathbf{EoR}$ signal extraction.
Since we are still using the fixed foreground residual $\mathbf{fg}_\text{fix}$, we assume that it has a similar effect on the $\mathbf{EoR}$ signal extraction results as in Section~\ref{fgnsEoR}.
We follow the case of the three observation times in the previous subsection, and the corresponding results are displayed in Fig.~\ref{diff_obs_cross_fg}.

We still consider observation times of 1752, 4380, and 13140 hours to compare with the results in the previous subsection. For the observation period of 1752 hours, 1200 epochs are necessary for the U-Net application. A 4380-hour observation period requires 1900 epochs, whereas a 13140-hour observation period demands 4000 epochs.
By comparing Fig.~\ref{diff_obs_cross} and Fig.~\ref{diff_obs_cross_fg}, we find that the results remain essentially the same for regions at higher $k_{\parallel}$ and lower $k_{\perp}$ (above the horizon line). Moreover, a significant change in the 2D coherence power spectrum is still observed when $k_{\perp}$ is $0.113$ for 1752 hours of observation. 
But for regions at lower $k_{\parallel}$ and higher $k_{\perp}$ (below the horizon line), Fig.~\ref{diff_obs_cross_fg} shows the same incoherence as Fig.~\ref{fix_ns_eor_ps} due to the power from the foreground.

\section{Summary and Conclusion}
\label{Conclusion}

Extracting the $\mathbf{EoR}$ signal from observations presents significant challenges due to inhomogeneous and spectrally varying $uv$-coverage, bright foreground emission, thermal noise, and various systematic effects such as beam errors. 
This paper examines SKA-Low-like observations in the SCP field, focusing on thermal noise and different systematic effects predominantly excess variance.
We assumed a similar $uv$ coverage as for LOFAR and restricted the analysis to the $50-250 \lambda$ baseline range, including the current flagging mask of LOFAR EoR observations of NCP filed. 
We concentrate on SKA-Low results, as LOFAR's noise levels are currently too high, whereas SKA's expected noise is approximately $5\%$ of LOFAR's current noise level.

We used foreground residuals from actual observations of the NCP field, excess power and thermal noise based on GPR, along with $\mathbf{EoR}$ signals produced by \textbf{\texttt{21cmFAST}}. 

We employed a 3D U-Net neural network for analyzing various sky maps. 
Initially, we evaluated a basic scenario that considered only the $\mathbf{ns}_\text{th}$ and $\mathbf{EoR}$ signals. 
Over 1752 hours of observation, the SKA's $\mathbf{ns}_\text{th}$ level is below that of the $\mathbf{EoR}$ signal, enabling U-Net to reliably produce a 2D power spectrum for $\mathbf{EoR}$.
We also examined the robustness of our results, showing that the 21-cm signal recovered in the wedge regions is not due to U-Net predictions from signals above the wedge.
The $30^\circ$ filter wedge was removed from the 2D power spectra of the $\mathbf{EoR_{ori}}$ signal, and the $\mathbf{EoR_{rev}}$ images were regenerated based on that. 
We also investigated the impact of $\mathbf{fg}_\text{fix}$ alongside previous findings, which revealed a significant discrepancy in the lower right of the 2D coherence power spectrum. 
This is consistent with the processing of real observations and is due to the dominance of the foreground in the large-scale structure of the line-of-sight direction.
Moreover, the joint influence of both $\mathbf{ns}_\text{th}$ and $\mathbf{ns}_\text{ex}$ on the $\mathbf{EoR}$ signal extraction were examined. Given that $\mathbf{ns}_\text{ex}$ exceeds the intensity of the $\mathbf{EoR}$ signal only after about 1752 hours, we approached the results with caution. 
Thus, observations were also extended to 4380 hours and 13140 hours. 
At 1752 hours, parts of the 2D coherence power spectrum above the horizon show weaker correlations, but regions with $k_{\perp}$ values below $0.113$ produced reliable results. 
For 4380 hours, reliable results were achieved within the entire EoR window. 
With 13140 hours, consistent 2D coherence power spectra were observed both above and below the horizon set by the excess variance. 
Lastly, considering the impact of $\mathbf{fg}_\text{fix}$ on the basis of $\mathbf{ns}_\text{th}$ and $\mathbf{ns}_\text{ex}$ only changes the signal extraction results from U-Net below the horizon line. 
Due to the power of the foreground, the lower right corner of the 2D coherence power spectrum again shows an inconsistency.
This suggests a promising future role for the 3D U-Net neural network in processing real SKA observational data, but that any presence of residual foreground and excess variance, as observed in current LOFAR data, could have a significant impact on the recovery of the 21-cm signal below the horizon delay line.  
Due to the incoherence of $\mathbf{ns}_\text{ex}$ in the frequency direction, these contaminations cannot be easily removed even with deep learning techniques and the most effective way forward is to reduce these contaminations by improving data calibrations and foreground subtraction methods as discussed by \cite{Acharya:2024qzn} and \cite{Mertens:2023dcl}.

\section*{Acknowledgements}

We are grateful for the support of the National SKA Program of China (Grants Nos. 2022SKA0110200 and 2022SKA0110203), the National Natural Science Foundation of China (Grant Nos. 12473001, 11975072, and
11875102), the Liaoning Revitalization Talents Program (Grant No. XLYC1905011), the 111 Project (Grant No. B16009), the Top-Notch Young Talents Program of China (Grant No. W02070050), the China Manned Space Project (Grant No. CMS-CSST-2025-A02), the European Research Council (ERC) under the European Union’s Horizon $2020$ research and innovation programme (Grant agreement No. 884760, "CoDEX”), Science and Engineering Research Board - Department of Science and Technology (SERB-DST) Ramanujan Fellowship (Grant agreement No. RJF/2022/000141)the European Research Council (ERC) under the European Union’s Horizon $2020$ research and innovation programme (Grant agreement No. 884760, "CoDEX”), Science and Engineering Research Board - Department of Science and Technology (SERB-DST) Ramanujan Fellowship (Grant agreement No. RJF/2022/000141), the Centre for Data Science and Systems Complexity (DSSC), Faculty of Science and Engineering at the University of Groningen, the Swedish Research Council (Grant agreement No. 2020-04691), and the Ministry of Universities and Research (MUR) through the PRIN project `Optimal inference from radio images of the epoch of reionization'.
YL acknowledges the support of the National Natural Science Foundation of China (Grant No. 12473091) and the Fundamental Research Funds for the Central Universities (Grant No. N2405008).

\bibliography{reference}{}
\bibliographystyle{aasjournal}

\appendix

\section{Subscripts and corresponding components}
\label{appendixA}

For ease of reading, we list the subscripts and corresponding components that appear in this paper in Table \ref{variables}.

\begin{table}
\centering
\caption{Subscripts and corresponding components.}
\begin{tabular}{ll}
\hline
$\text{fg}$&component of foregrounds\\
$\text{ex}$&component of excess variances\\
$\text{th}$&component of thermal noises \\
$\text{EoR}$&component of 21-cm signals from Epoch of Reionization \\
$\text{int}$&component of intrinsic sky emissions \\
$\text{mix}$&component of mode-mixing contaminants \\
$\text{ori}$&component of the sky maps obtained by inverse Fourier transform before using a filter wedge with a $30^\circ$ angle in the 2D power spectra \\
$\text{rev}$&component of the sky maps obtained by inverse Fourier transform after using a filter wedge with a $30^\circ$ angle in the 2D power spectra \\
$\text{fix}$&component of fixed smooth foreground residual \\
$\text{all}$&component of thermal noises and excess variances added together \\
\hline
\end{tabular}
\centering
\label{variables}
\end{table}

\section{LOFAR results with $\mathbf{ns}_\text{th}$ + $\mathbf{EoR}$}
\label{appendixB}

\begin{figure*}[h]
\centering
\includegraphics[width=0.90\textwidth]{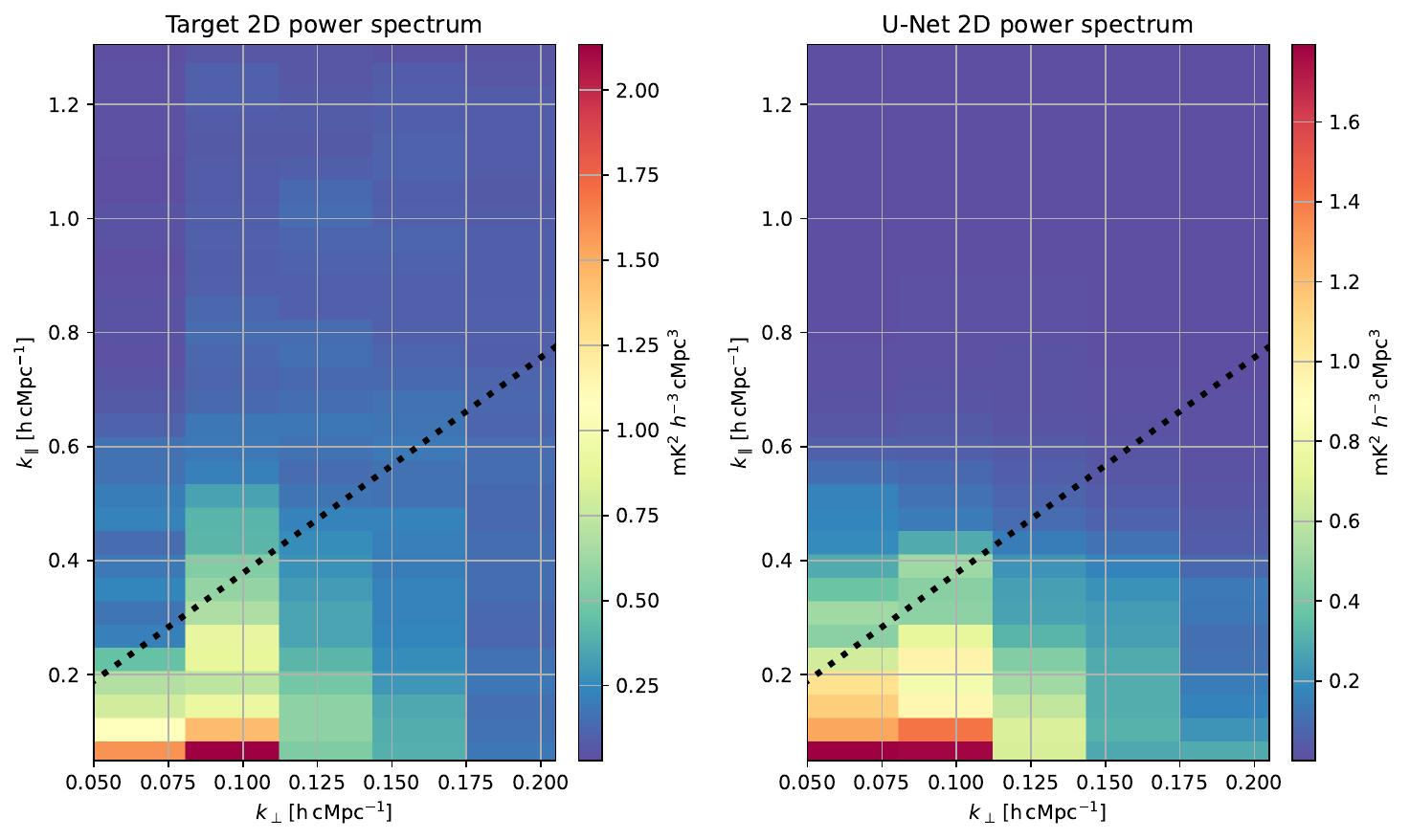}
\caption{\label{LOFAR_ns_eor_ps}2D power spectrum of the target $\mathbf{EoR}$ image and the 2D power spectrum of the predicted image given by U-Net after 850 epochs for LOFAR scenario when only consider $\mathbf{ns}_\text{th}$. The black dotted lines are horizon lines for LOFAR.}
\end{figure*}

\begin{figure*}[h]
\centering
\includegraphics[width=0.90\textwidth]{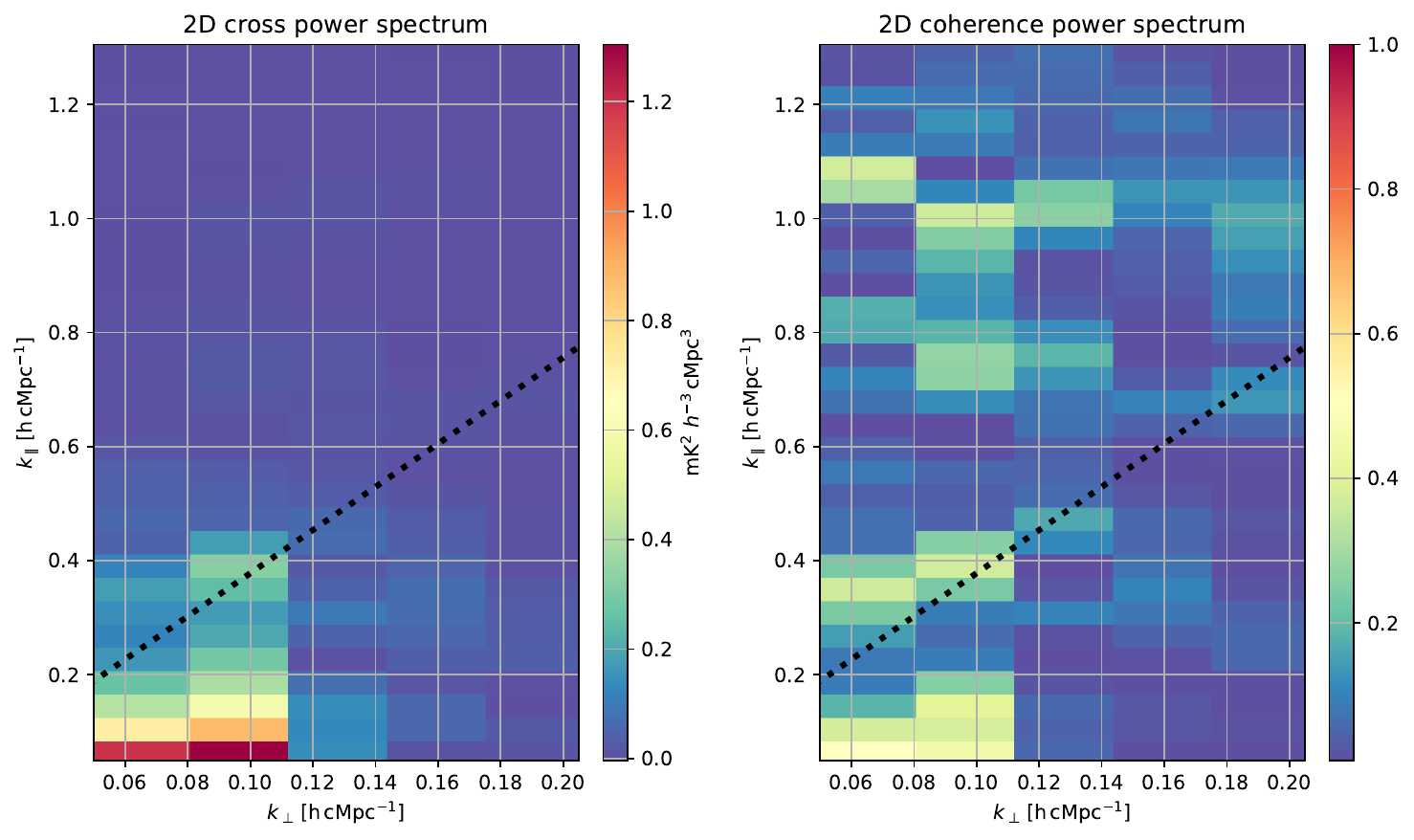}
\caption{\label{LOFAR_ns_eor_cps}2D cross power spectrum and 2D coherence power spectrum of the target and predicted images for LOFAR scenario when only consider $\mathbf{ns}_\text{th}$. The black dotted lines are horizon lines for LOFAR.}
\end{figure*}

Here we test the results of 1752 hours of LOFAR observations considering only the effect of $\mathbf{ns}_\text{th}$.
U-Net reaches optimal loss after 850 epochs.
The 2D power spectra of the target $\mathbf{EoR}$ and the U-Net prediction result are shown on the left and right of Fig.~\ref{LOFAR_ns_eor_ps}, respectively.
Their corresponding 2D cross power spectrum and 2D coherence power spectrum are displayed in Fig.~\ref{LOFAR_ns_eor_cps}.
Although the target power spectrum and the power spectrum of the U-Net prediction result have similar shapes, they differ by an order of magnitude and have significant inconsistencies in the coherent power spectrum.

\section{LOFAR results with $\mathbf{ns}_\text{ex}$ + $\mathbf{ns}_\text{th}$ + $\mathbf{EoR}$}
\label{appendixC}

\begin{figure*}
\centering
\includegraphics[width=0.90\textwidth]{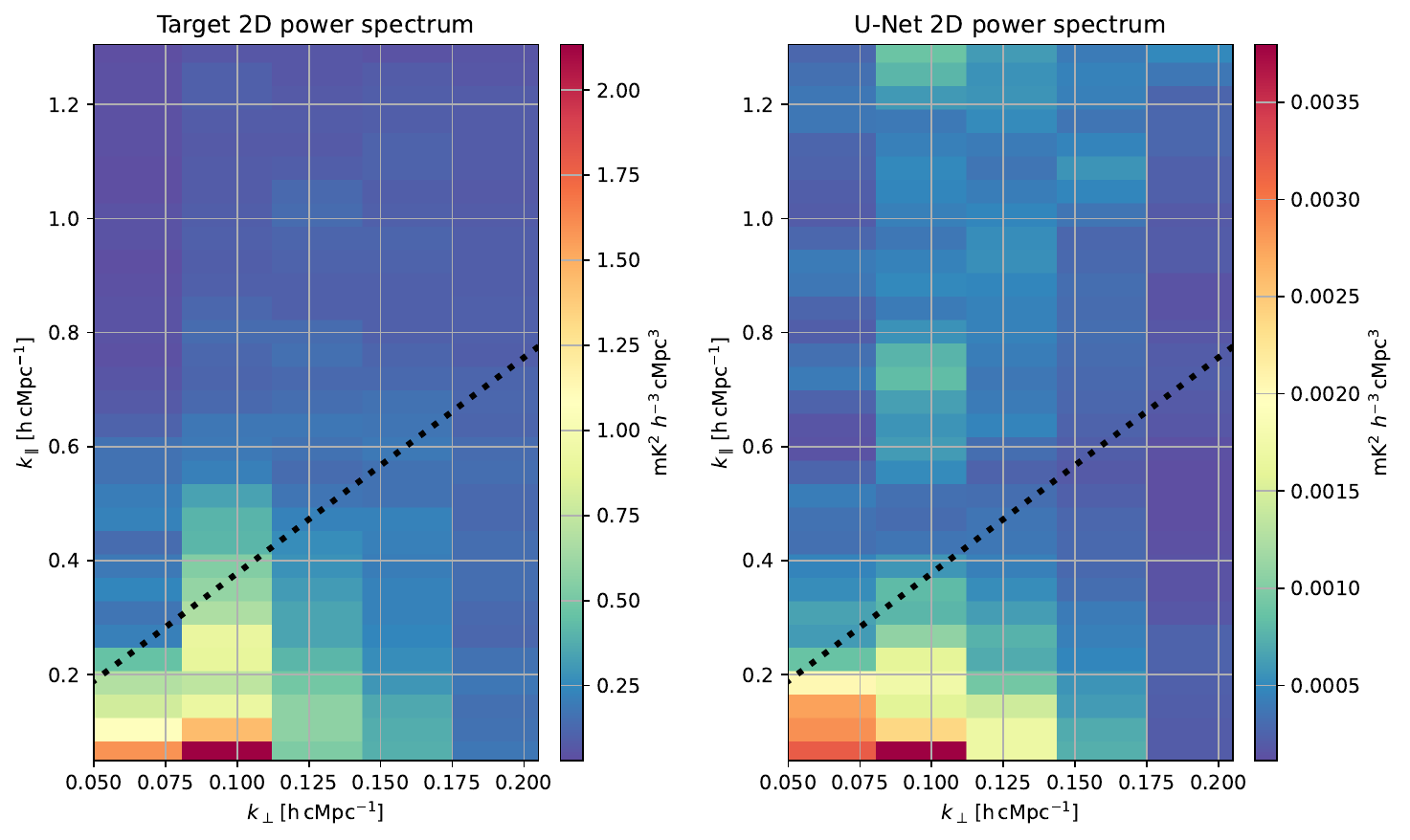}
\caption{\label{LOFAR_ex_ns_eor_ps}2D power spectrum of the target $\mathbf{EoR}$ image and the 2D power spectrum of the predicted image given by U-Net after 1500 epochs for LOFAR scenario when considering $\mathbf{ns}_\text{th}$ and $\mathbf{ns}_\text{ex}$. The black dotted lines are horizon lines for LOFAR.}
\end{figure*}

\begin{figure*}
\centering
\includegraphics[width=0.90\textwidth]{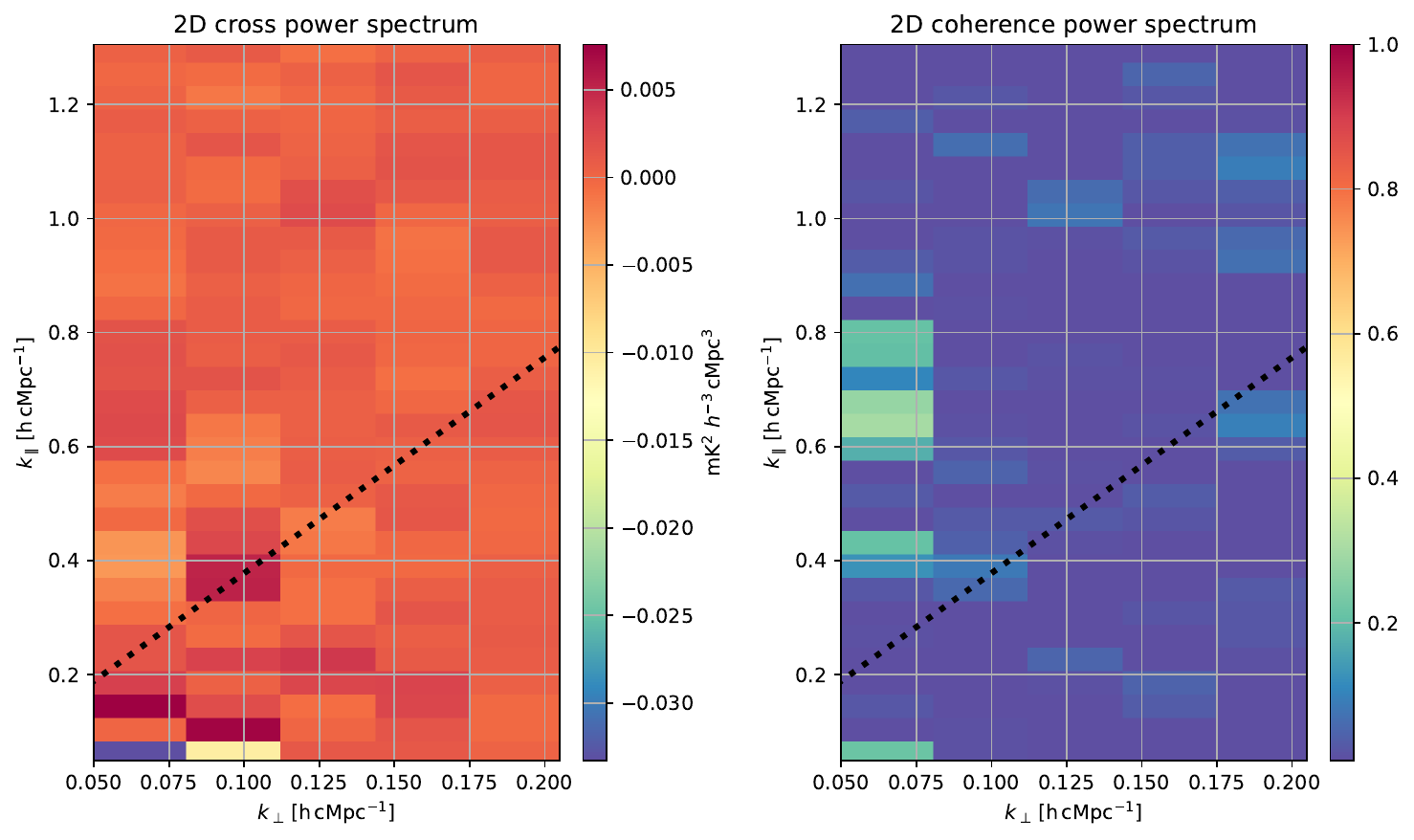}
\caption{\label{LOFAR_ex_ns_eor_cps}2D cross power spectrum and 2D coherence power spectrum of the target and predicted images for LOFAR scenario when considering $\mathbf{ns}_\text{th}$ and $\mathbf{ns}_\text{ex}$. The black dotted lines are horizon lines for LOFAR.}
\end{figure*}

Here we add to the above the effect of $\mathbf{ns}_\text{ex}$ with 1752 hours of LOFAR observations.
Due to the high intensity of $\mathbf{ns}_\text{th}$ and $\mathbf{ns}_\text{ex}$, U-Net could not learn small-scale structures, so it reached the minimum loss after 1500 epochs.
Similarly, Fig.~\ref{LOFAR_ex_ns_eor_ps} illustrates the target 2D power spectrum and the 2D power spectrum of the U-Net results, and Fig.~\ref{LOFAR_ex_ns_eor_cps} shows their 2D cross power spectrum and 2D coherence power spectrum.
Undoubtedly we get worse results, and on the 2D coherence power spectrum you can see that there is almost no correlation between the two images.

\end{document}